\documentclass[12pt,draftclsnofoot,letterpaper,oneside,onecolumn]{IEEEtran}
\def\BibTeX{{\rm B\kern-.05em{\sc i\kern-.025em b}\kern-.08em
    T\kern-.1667em\lower.7ex\hbox{E}\kern-.125emX}}
\usepackage{amsmath}
\usepackage{amsthm}
\usepackage{amsfonts}
\usepackage{amssymb}
\usepackage{graphicx}
\usepackage{cite}
\usepackage{subfigure}
\usepackage{balance}
\usepackage{algorithm}
\usepackage{algorithmic}
\usepackage{enumerate}
\usepackage{color}
\usepackage{array}
\usepackage{indentfirst}
\usepackage{multirow}
\usepackage{booktabs}
\usepackage{color}
\usepackage{slashbox}
\usepackage{diagbox}
\usepackage{threeparttable}
\usepackage[svgnames,table]{xcolor}
\usepackage{bm}
\usepackage{bbm}
\usepackage{subfigure}
\usepackage{epstopdf}

\makeatletter

\newcommand{\Rmnum}[1]{\expandafter\@slowromancap\romannumeral #1@}
\makeatother

\begin{document}
 	\title{{Predictive Precoder Design for OTFS-Enabled URLLC: A Deep Learning Approach}}

\author{\IEEEauthorblockN{Chang Liu, \emph{Member, IEEE}, Shuangyang Li, \emph{Member, IEEE}, Weijie Yuan, \emph{Member, IEEE}, Xuemeng Liu, and Derrick Wing Kwan Ng, \emph{Fellow, IEEE} \vspace{-0.8cm}}

\thanks{C. Liu and D. W. K. Ng are with the University of New South Wales, Sydney, Australia. S. Li is with the University of Western Australia, Perth, Australia. W. Yuan is with the Southern University of Science and Technology, Shenzhen, China. X. Liu is with the University of Sydney, Sydney, Australia.}

}

\maketitle

\begin{abstract}
This paper investigates the orthogonal time frequency space (OTFS) transmission for enabling ultra-reliable low-latency communications (URLLC).
To guarantee excellent reliability performance, pragmatic precoder design is an effective and indispensable solution.
However, the design requires accurate instantaneous channel state information at the transmitter (ICSIT) which is not always available in practice.
Motivated by this, we adopt a deep learning (DL) approach to exploit implicit features from estimated historical delay-Doppler domain channels (DDCs) to directly predict the precoder to be adopted in the next time frame for minimizing the frame error rate (FER), that can further improve the system reliability without the acquisition of ICSIT.
To this end, we first establish a predictive transmission protocol and formulate a general problem for the precoder design where a closed-form theoretical FER expression is derived serving as the objective function to characterize the system reliability.
Then, we propose a DL-based predictive precoder design framework which exploits an unsupervised learning mechanism to improve the practicability of the proposed scheme.
As a realization of the proposed framework, we design a DDCs-aware convolutional long short-term memory (CLSTM) network for the precoder design, where both the convolutional neural network and LSTM modules are adopted to facilitate the spatial-temporal feature extraction from the estimated historical DDCs to further enhance the precoder performance.
Simulation results demonstrate that the proposed scheme facilitates a flexible reliability-latency tradeoff and achieves an excellent FER performance that approaches the lower bound obtained by a genie-aided benchmark requiring perfect ICSI at both the transmitter and receiver.

\vspace{-0.2cm}
\end{abstract}

\begin{IEEEkeywords}\vspace{-0.2cm}
Ultra-reliable low-latency communications (URLLC), orthogonal time frequency space (OTFS), predictive precoder design, deep learning.
\end{IEEEkeywords}

\clearpage

\section{Introduction}
Ultra-reliable and low-latency communications (URLLC) is one of the key pillars of the current fifth-generation (5G) wireless networks and is envisioned as a disruptive technology for the coming sixth-generation (6G)~\cite{park2022extreme, She2017commag, wong2017key}. Numerous emerging applications fall into the use cases of URLLC, including the automotive \cite{dar2010wireless}, Industry 4.0 \cite{siqin2022platform}, and emergency relief~\cite{popovski2019wireless}.
In practice, the key features for URLLC are the ultra-high reliability and low latency, where the block error rate is required to be lower than $0.001 \%$ and the latency is shorter than 1 milliseconds (ms)~\cite{park2022extreme}.

To satisfy these stringent requirements of URLLC, various designs have been developed in the literature.
For instance, a cross-layer optimization method for URLLC was proposed in~\cite{She2018cross_layer}, where both the transmission and queueing delays were considered. Specifically, a proactive packet dropping mechanism was introduced based on which the packet dropping policy, power allocation policy, and bandwidth allocation policy were then optimized for the transmit power minimization taking into account the quality of service (QoS) constraint of users. Besides, in~\cite{Chang2019optimizing}, a joint resource optimization for both URLLC and control subsystems was considered to maximize the overall system performance. In particular, the authors firstly converted the control convergence rate requirement into a communication reliability constraint. Then, an iterative algorithm was proposed to acquire the optimal solution to the resource allocation problem where numerical results have demonstrated a remarkable performance gain in terms of the spectral efficiency and control cost.
Also, a tutorial on URLLC was presented in~\cite{she2021Tutorial}, where the state-of-art designs for URLLC based on deep learning was provided. Specifically, a detailed illustration on how domain knowledge of communications and networking can be integrated into different kinds of deep learning algorithms for URLLC systems was provided.
On the other hand, long channel codes are generally capable to provide substantial coding gain to compact channel impairments. However, a major challenge for URLLC in future wireless communications is the low latency constraint that limits the use of long blocklength. The short block length inevitably limits the coding gain and leads to a degraded decoding error performance comparing to the counterparts adopting a long channel code.
For example, the celebrating orthogonal frequency-division multiplexing (OFDM) \cite{farhang2016ofdm} requires the application of complex channel encoding across several OFDM symbols to ensure the required channel diversity. However, the use of short block length greatly limits the achievable coding gain such that OFDM would suffer from severe diversity loss (due to the insufficient Hamming distance between codeword pairs), resulting in an unsatisfactory reliability \cite{wong2017key, sutton2019enabling}.

In contrast to the conventional time-frequency (TF) domain signal processing in OFDM, orthogonal time frequency space (OTFS) modulation \cite{Hadani2017orthogonal, Zhiqiang_magzine} relies on the signal processing in the delay-Doppler (DD) domain with quasi-static, sparse, and compact properties, which positions OTFS as a promising solution to the realization of URLLC.
Furthermore, since the information spreading performed by OTFS ensures that each DD domain information symbol experiences the whole channel fluctuation in the TF domain, OTFS has the potential of achieving the full channel diversity, which is necessary for the ultra-high reliability required by URLLC.
For example, the diversity of uncoded OTFS transmission was studied in \cite{Raviteja2019effective}, which indicates that OTFS can almost achieve the full channel diversity and the probability of OTFS having full channel diversity increases with the frame size. Also, in~\cite{Li2020performance}, the error performance of coded OTFS was reported and the authors unveiled that there is a non-trivial tradeoff between the diversity gain and the coding gain in OTFS transmission, where the diversity gain increases with the number of independent resolvable paths, while the coding gain declines.
It should be noted that most studies on OTFS error performance relied on the complicated maximum-likelihood detection, that introduces a prohibitively high detection complexity, especially when the channel is with large numbers of resolvable paths as expected in  6G~\cite{Raviteja2018interference}.
As an alternative, more practical detection methods can be adopted, such as the zero-forcing (ZF) and minimum mean square error (MMSE) detectors \cite{tse2005fundamentals, kay1993fundamentals}, which however, only achieve limited detection performance.
To address this issue, a precoded OTFS transmission scheme was proposed in~\cite{HuangQin2022ComLet}, where the precoder was optimized to minimize the system bit error rate. However, such a precoder relies on the perfect knowledge of instantaneous channel state information at the transmitter (ICSIT), which is highly impractical in reality. Consequently, precoder designs based on the \emph{outdated} ICSIT inevitably degrade the error performance.

Based on the above discussion, the acquisition of ICSIT hinders the design of a practical precoder for URLLC.
It is worth noting that if we can predict the desired precoder in advance, we can not only bypass the acquisition of ICSIT but also guarantee the system reliability in the subsequent period.
Meanwhile, the deep learning (DL) technology \cite{goodfellow2016deep} has proven its efficiency and effectiveness in various wireless applications \cite{wang2017deep}, such as signal detection \cite{liu2020deeptransfer} and classification \cite{hu2019deep, xie2020deep}, channel estimation \cite{lxm2020deepresidual, liu2020deepresidualconference}, resource allocation \cite{liu2022learning}, etc. 
In particular, the recurrent neural network (RNN) \cite{goodfellow2016deep, hochreiter1997long} can exploit itself recurrent feedback connections to exhibit the temporal dynamic behavior, thus empowering RNNs with a powerful capability in extracting temporal features from the input data to facilitate the predictive tasks.
Motivated by this, in this paper, we adopt an RNN-based approach to implicitly learn features of subsequent channels from the historical channels to predict the precoder to be adopted in the subsequent period, which further improves the system reliability without the availability of ICSIT.
The main contributions of our work are summarized as follows:
\begin{enumerate}[(1)]
\item We develop a predictive transmission protocol and formulate a universal problem for predictive precoder design to maximize the system reliability subject to the transmit power constraint.
    In the formulated problem, we first derive a closed-form frame error rate (FER) expression theoretically taking into account both the precoder at the transmitter and the equalizer at the  receiver to characterize the system reliability.
    In particular, our formulated problem does not require the availability of ICSIT and only exploits estimated historical channels to predict precoder to be adopted in the next time frame, which provides a more practical precoder design scheme.

\item To address the formulated precoder design problem, we develop a versatile DL-based predictive precoder framework for OTFS-enabled URLLC, which consists of offline unsupervised training and online precoder design.
   Specifically, in the developed framework, we exploit an unsupervised learning mechanism to further improve the practicability of the adopted DL approach.

\item To provide a realization of the developed framework, a DD domain channels-aware convolutional long short-term memory (CLSTM) \cite{shi2015convolutional} network (DDCL-Net) is designed for predictive precoder design in OTFS-enabled URLLC systems.
    In the designed DDCL-Net, both convolutional neural network (CNN) \cite{krizhevsky2012imagenet} and LSTM \cite{hochreiter1997long} modules are adopted to jointly exploit spatial-temporal features from historical communication channels to improve the designed neural network performance for precoder design.

\item Extensive simulation results have been presented to verify the effectiveness and efficiency of the proposed scheme in terms of FER performance, generalizability, and reliability-latency tradeoff.
    In particular, our results show that the FER performance of the proposed predictive scheme approaches to that of the lower bound obtained by the DL-based benchmark method with the availability of perfect ICSI at both the transmitter and the receiver.

\end{enumerate}

The remainder of this paper is organized as follows.
In Section \Rmnum{2}, we introduce the system model of the considered OTFS-enabled URLLC.
Then, Section \Rmnum{3} develops a predictive transmission protocol and formulates a universal problem for the considered predictive precoder design.
In Section \Rmnum{4} and Section \Rmnum{5}, we propose a versatile DL-based precoder design framework and design a DDCL-Net-based precoder as a realization of the developed framework.
Subsequently, extensive simulations are conducted to evaluate the performance of the developed scheme in Section \Rmnum{6}.
Section \Rmnum{7} finally provides conclusions for this paper.

\emph{Notations:}
We use superscripts $H$ and $T$ to denote the conjugate transpose and transpose, respectively.
$\mathbb{R}$ and $\mathbb{C}$ denote the sets of real and complex numbers, respectively.
$\mathbb{N}_1$ represents the natural number set.
$ \otimes $ denotes the kronecker product.
${{{\bf{F}}_N}}$ and ${{{\bf{I}}_M}}$ denote the discrete Fourier transform (DFT) matrix of size $N\times N$ and the identity matrix of size $M\times M$, respectively.
$\textrm{vec}(\cdot)$ denotes the vectorization operation and $[a]_b$ denotes the modulo operation of $a$ modulo $b$.
$\textrm{diag}(\mathbf{a})$ denotes the generation of a diagonal matrix with the input vector $\mathbf{a}$ as the diagonal elements.
${\mathcal{CN}}( \bm{\mu},\mathbf{\Sigma} )$ denotes a circularly symmetric complex Gaussian (CSCG) distribution with $\bm{\mu}$ and $\mathbf{\Sigma}$ being the associated statistical mean vector and covariance matrix, respectively.
$U(a,b)$ denotes a uniform distribution with the range of $[a,b]$.
$|\cdot|$ denotes the absolute value of a complex-valued number.
$\|\cdot\|$ and $\|\cdot\|_F$ represent the $\ell_2$ and Frobenius norms of a vector and a matrix, respectively.
$\sqrt{x}$ denotes the square root operation on $x$.
$\Pi_{k=1}^K x_k$ denotes the product of $K$ numbers $x_1,x_2,\cdots, x_K$.
In addition, we adopt $\mathbf{X}(a:b,:)$ to denote the sub-matrix generated from the $a$-th row to the $b$-th row of matrix $\mathbf{X}$.
$\mathbb{E}\{\cdot\}$ refers to the calculation of statistical expectation.
$\mathrm{Re}\{\cdot\}$ and $\mathrm{Im}\{\cdot\}$ are adopted to denote the real and imaginary parts of a complex-valued scalar/vector/matrix, respectively.

\begin{figure}
\centering
\includegraphics[width=0.8\textwidth]{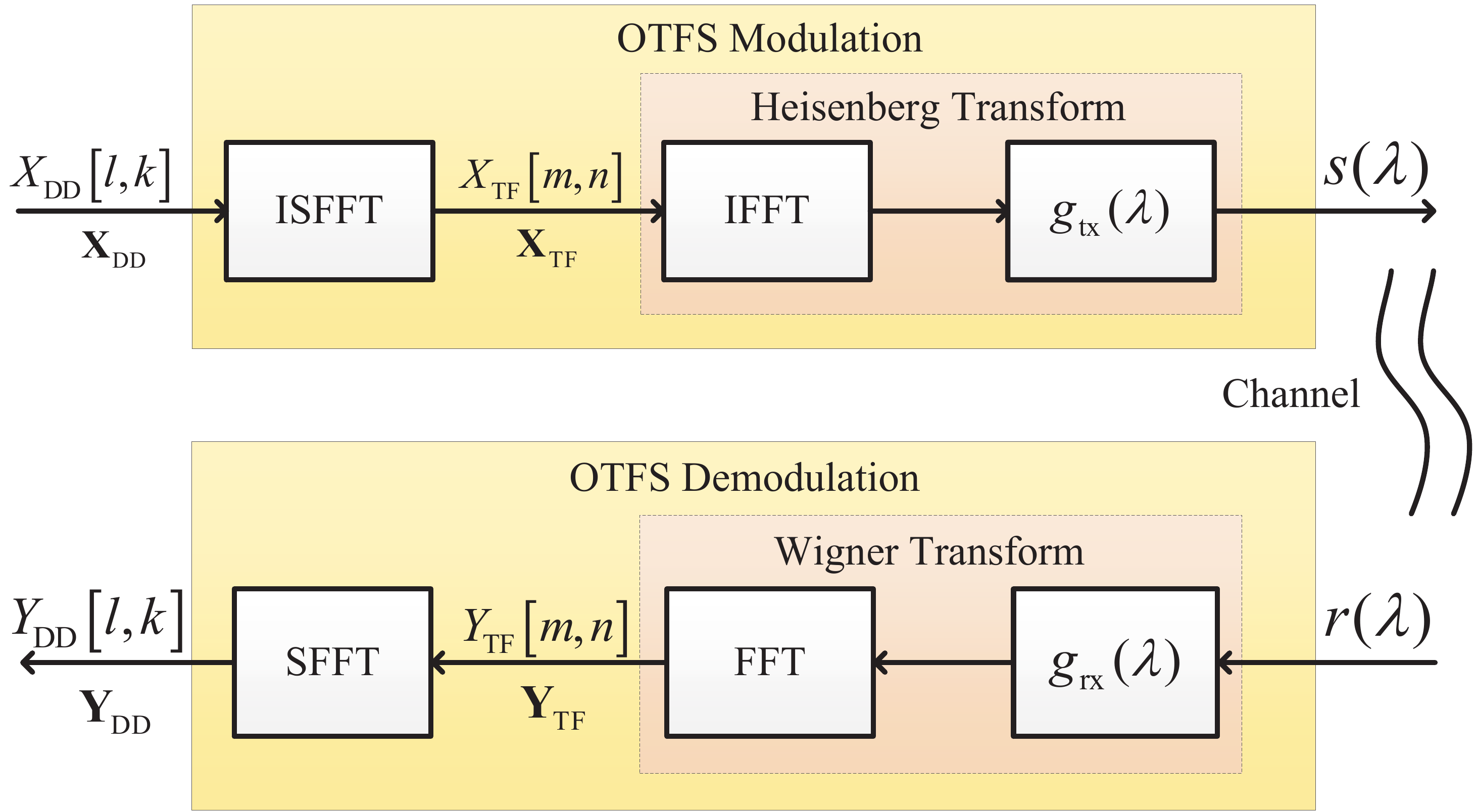}
\caption{The considered OTFS-enabled URLLC system.}
\label{scenario}
\centering
\end{figure}

\section{System Model}
\subsection{OTFS System Model}
Without loss of generality, we consider an OTFS-enabled URLLC system as shown in Fig. \ref{scenario}, where both the transmitter and receiver are single-antenna devices.
Denote by $M$ the number of delay bins/number of subcarriers and $N$ the number of Doppler bins/number of time slots, respectively.
In addition, each time slot duration is $T$ and the subcarrier spacing is $1/T$.
Let ${\bf X}_{\rm DD} \in \mathbb{C}^{M \times N} $ be the DD domain symbol matrix.
In this case, the $(l,k)$-th, $l \in \{0,1,\cdots,M-1\}$, $k \in \{0,1,\cdots,N-1\}$, entry of ${\bf X}_{\rm DD} $, denoted by ${X_{{\rm{DD}}}}\left[ {l,k} \right]$, is the $(l,k)$-th DD domain symbol, which is uniformly generated from an energy-normalized constellation set $\mathbb A$.
Similarly, let ${\bf X}_{\rm TF} \in \mathbb{C}^{M \times N}$ be the TF domain symbol matrix, thus the $(m,n)$-th, $m \in \{0,1,\cdots,M-1\}$, $n \in \{0,1,\cdots,N-1\}$, element of ${\bf X}_{\rm TF}$, denoted by $X_{\rm TF}[ {m,n} ]$ is the $(m,n)$-th TF domain symbol.
According to \cite{Hadani2017orthogonal}, by exploiting the inverse symplectic finite Fourier transform (ISFFT), $X_{\rm TF}\left[ {m,n} \right]$ can be formulated as
\begin{equation}
X_{\rm TF}\left[ {m,n} \right] = \frac{1}{{\sqrt {MN} }}\sum\limits_{k = 0}^{N - 1} {\sum\limits_{l = 0}^{M - 1} {X_{\rm DD}\left[ {l,k} \right]} } {e^{j2\pi \left( {\frac{{nk}}{N} - \frac{{ml}}{M}} \right)}}  . \label{ISFFT}
\end{equation}
As depicted in Fig. \ref{scenario}, to obtain the OTFS signal at time instant $\lambda \in \mathbb{R}$, denoted by $s(\lambda)$, we can perform the Heisenberg transform \cite{Hadani2017orthogonal} on ${\bf X}_{\rm TF}$ based on the shaping pulse $g_{\rm tx}\left( \lambda \right)$.
In particular, the Heisenberg transform is commonly implemented by the inverse fast Fourier transform (IFFT), i.e., the conventional OFDM modulation. Thus, the resultant OTFS signal $s\left( \lambda \right)$ can be formulated as
\begin{equation}
s( \lambda ) = \sum\limits_{n = 0}^{N - 1} {\sum\limits_{m = 0}^{M - 1} {X_{\rm TF}\left[ {m,n} \right]{g_{{\rm{tx}}}}\left( {\lambda - nT} \right){e^{j2\pi m\Delta f\left( {\lambda - nT} \right)}}} },\label{OTFS_signal}
\end{equation}
where $\Delta f$ denotes frequency spacing between adjacent subcarriers.
For ease of study, we assume that the OTFS signal $s(\lambda)$ is transmitted over a DD deterministic channel as commonly adopted in e.g., \cite{li2021cross, shen2019channel, raviteja2019embedded}, where the delay and Doppler responses of the underlying physical scatterers remain static over the signal transmission.
With this assumption, a general linearly time variant (LTV) channel with $P$ resolvable paths can be modeled as~\cite{Hadani2017orthogonal, hlawatsch2011wireless}
\begin{equation}
{h_{{\rm{DD}}}}\left( {\tau ,\nu } \right) =  \sum\limits_{p = 1}^P {{h_p}\delta \left( {\tau  - {\tau _p}} \right)} \delta \left( {\nu  - {\nu _p}} \right),\label{DD_channel}
\end{equation}
where $h_p$, $\tau_p$, and $\nu_p$ are the complex fading coefficient, the time delay, and the Doppler frequency associated to the $p$-th resolvable path, for $1 \le p \le P$, respectively.
Let $r\left( \lambda \right)$ be the time domain received signal which is given by~\cite{Hadani2017orthogonal}
\begin{align}
r\left( \lambda \right) = \int_{ - \infty }^\infty  {\int_{ - \infty }^\infty  {{h_{{\rm{DD}}}}\left( {\tau ,\nu } \right)} } {e^{j2\pi \nu \left( {\lambda - \tau } \right)}}s(\lambda){\rm{d}}\nu {\rm{d}}\tau + w(\lambda),
\label{TD_twisted_conv_delay_first_Doppler_second}
\end{align}
where $w(\lambda) \sim \mathcal{CN}(0,\sigma^2)$ denotes the CSCG noise \cite{liu2019maximum} with a zero-mean and variance $\sigma^2$. As indicated by Fig.~\ref{scenario}, the OTFS demodulator can be represented by the Wigner transform with the matched-filtering pulse $g_{\rm rx}\left( \lambda \right)$ followed by the SFFT. Corresponding to the transmitter side, the Wigner transform can be implemented by the conventional OFDM demodulation,
Thus, we can obtain the received TF domain symbol ${Y_{{\rm{TF}}}}\left[ {m,n} \right]$, which is given by~\cite{Raviteja2018interference}
\begin{align}
{Y_{{\rm{TF}}}}\left[ {m,n} \right] = \int_{0}^{NT} {r\left( {\lambda'} \right)} g_{{\rm{rx}}}^*\left( {\lambda' - nT} \right){e^{ - j2\pi m\Delta f\left( {\lambda' - nT} \right)}}{\rm{d}}\lambda'.
\label{TF_receied}
\end{align}
Correspondingly, the received DD domain symbol ${Y_{{\rm{DD}}}}\left[ {l,k} \right]$ can be obtained by applying the SFFT and it can be formulated as~\cite{Hadani2017orthogonal}
\begin{align}
{Y_{{\rm{DD}}}}\left[ {l,k} \right] = \frac{1}{{\sqrt {MN} }}\sum\limits_{n = 0}^{N - 1} {\sum\limits_{m = 0}^{M - 1} {{Y_{{\rm{TF}}}}\left[ {m,n} \right]{e^{ - j2\pi \left( {\frac{{nk}}{N} - \frac{{ml}}{M}} \right)}}} } .\label{Y_DD}
\end{align}

For ease of presentation, let us focus on the corresponding matrix representation of the above derivations in the following. Without loss of generality, we consider the popular rectangular pulse for transmit pulse shaping and receive matched-filtering. Thus, according to~\cite{Raviteja2019practical}, the discrete time domain OTFS symbol vector $\bf s$ of length $MN$ corresponding to~\eqref{OTFS_signal} can be written as
\begin{align}
{\bf{s}} = \left( {{\bf{F}}_N^{\rm{H}} \otimes {{\bf{I}}_M}} \right){{\bf{x}}_{{\rm{DD}}}},
\end{align}
where ${{\bf{x}}_{{\rm{DD}}}} \buildrel \Delta \over = {\rm{vec}}\left( {{{\bf{X}}_{{\rm{DD}}}}} \right) \in \mathbb{C}^{MN \times 1}$ denotes the vectorized DD domain information symbols.
According to \cite{Raviteja2019practical}, the delay resolution is the inverse of the total bandwidth $T/M$, while the Doppler resolution is the inverse of the frame duration $1/(NT)$. Then, denote by ${\tau _p}$ and ${\nu _p}$ the delay and Doppler shift, respectively, we have
\begin{align}
{\tau _p} = \left( {{l_p} + {\iota _p}} \right)\frac{T}{M}\label{delay_indices} ,
\end{align}
and
\begin{align}
{\nu _p} = \frac{{{k_p} + {\kappa _p}}}{{NT}}\label{Doppler_indices},
\end{align}
respectively.
Here, ${l_p}$ and ${\iota _p}$ represent the integer and fractional delay indices, respectively, and ${k_p}$ and ${\kappa_p}$ denote the integer and fractional Doppler indices, respectively.
Note that the fractional delay and Doppler indices satisfy both $ - {1 \mathord{\left/
 {\vphantom {1 2}} \right.
 \kern-\nulldelimiterspace} 2} \le{\iota _p}\le {1 \mathord{\left/
 {\vphantom {1 2}} \right.
 \kern-\nulldelimiterspace} 2}$ and $- {1 \mathord{\left/
 {\vphantom {1 2}} \right.
 \kern-\nulldelimiterspace} 2} \le {\kappa _p} \le {1 \mathord{\left/
 {\vphantom {1 2}} \right.
 \kern-\nulldelimiterspace} 2}$, which correspond to the fractional shifts from the nearest delay and Doppler indices~\cite{Raviteja2018interference}, respectively.
Generally, the typical value of the sampling time $1/(M\Delta f)$ in the delay domain is sufficiently small.
Therefore, the impact of fractional delays in typical wideband systems can be neglected~\cite{tse2005fundamentals}.
In this case, the received discrete time domain symbol vector (after CP removal) under the reduced-CP format can be expressed as~\cite{Raviteja2019practical}
\begin{align}
r_i  = \sum\limits_{p = 1}^P {{h_p}{e^{  j2\pi \frac{{\left( {{k_p} + {\kappa _p}} \right)\left( {i - {l_p}} \right)}}{{MN}}}}s_{{{\left[ {i - {l_p}} \right]}_{MN}}}  + w_i} , \label{TD_model_symbol_wise}
\end{align}
where subscript $i$ denotes the $i$-th, $i \in \{0,1,\cdots,MN-1\}$, sample.
Similar to~\cite{Raviteja2019practical}, by introducing the permutation matrix (forward cyclic shift)
\begin{equation}
{{\bf{ \Pi }}}= {\left[ {\begin{array}{*{20}{c}}
0& \cdots &0&1\\
1& \ddots &0&0\\
 \vdots & \ddots & \ddots & \vdots \\
0& \cdots &1&0
\end{array}}
\right]} \in \mathbb{R}^{MN \times MN}
\end{equation}
and the phase rotating matrix ${\bm \Delta}={\rm{diag}}( {1,{e^{j2\pi \frac{1}{{MN}}}},...,{e^{j2\pi \frac{{MN - 1}}{{MN}}}}} ) \in \mathbb{C}^{MN \times MN}$, we can write \eqref{TD_model_symbol_wise} into a vector form, that yields
\begin{align}
{\bf{r}} = {{\bf{H}}_{\rm{T}}}{\bf{s}} + {\bf{w}}, \label{TD_model_vec}
\end{align}
where ${\bf{r}} \in \mathbb{C}^{MN \times 1}$ is the received time domain symbol vector, $\mathbf{w} \sim \mathcal{CN}(\mathbf{0},\sigma^2\mathbf{I}_{MN})$ is the noise vector, and
\begin{align}
{{\bf{H}}_{\rm{T}}} \buildrel \Delta \over = \sum\limits_{p = 1}^P {{h_p}{e^{ - j2\pi \frac{{\left( {{k_p} + {\kappa _p}} \right){l_p}}}{{MN}}}}{{\bm \Delta} ^{{k_p} + {\kappa _p}}}{{\bf{\Pi }}^{{l_p}}}} \label{H_T}
\end{align}
is the effective time domain channel matrix.

\subsection{Channel Offset Model in OTFS-enabled URLLC}
In this section, we adopt a channel offset model to characterize the time-varying wireless environment.
In the considered OTFS-enabled URLLC, each data frame includes a short package where there is an $MN$-length symbol sequence to be transmitted, as introduced in Section A.
In this case, we assume that ${{\bf{H}}_{\rm{T}}}$ varies with frames and remains constant within each frame.
Note that duration of DD domain channel is characterized by the so-called ``stationarity region'', which is typically much larger than the coherence region in the TF domain \cite{Hadani2017orthogonal, Zhiqiang_magzine}. However, in the transmission of mobile users, the underlying channel geometry changes over frames according to the physical scatterers in the channel \cite{tse2005fundamentals}. Therefore, we adopt a channel offset model to characterize the delay and Doppler indices, which are given by \cite{liu2020location, yuan2020learning, liu2021deeplearningconference} 
\begin{equation}\label{}
  l_{p,t} = l_{p,t-1} + \epsilon_{p,t}
\end{equation}
and
\begin{equation}\label{}
  \tilde{k}_{p,t} = \tilde{k}_{p,t-1} + \varepsilon_{p,t}.
\end{equation}
Here, $l_{p,t}$ and $\epsilon_{p,t}$ denote the integer delay index and delay index-offset at frame $t$ associated to the $p$-th resolvable path, respectively.
$\forall p,t$, $\epsilon_{p,t} \sim U(\epsilon_{\min},\epsilon_{\max})$, where $\epsilon_{\min}$ and $\epsilon_{\max}$ denote the minimum and maximum delay index-offsets, respectively.
Similarly, $\tilde{k}_{p,t} = k_{p,t} + \kappa_{p,t}$ and $\varepsilon_{p,t} \sim U(\varepsilon_{\min},\varepsilon_{\max})$ denote the Doppler index and Doppler index-offset at frame $t$ associated to the $p$-th resolvable path, respectively, where $\varepsilon_{\min}$ and $\varepsilon_{\max}$ denote the minimum and maximum Doppler index-shifts, respectively.
In addition, without loss of generality, we adopt a first-order complex Gauss-Markov model \cite{nasir2019multi, xie2019activity, xie2020unsupervised} to characterize the time-varying nature of channel fading, i.e.,
\begin{equation}\label{}
  h_{p,t} = \rho h_{p,t-1} + \sqrt{1 - \rho^2} \vartheta_{p,t}.
\end{equation}
Here, $h_{p,t}$ is the fading coefficient and at frame $t$ associated to the $p$-th resolvable path.
$\rho \in [0,1]$ is the correlation coefficient.
$\forall p,t$, $\vartheta_{p,t} \sim \mathcal{CN}(0,\sigma_h^2)$ is the independent and identically distributed (i.i.d.) CSCG random variable \cite{liu2014maximum} with $\sigma_h^2 = 1/P$ \cite{li2021cross, liu2016blind}.

Based on above discussion, we can redefine the signal model by taking into consideration the time-varying property. In this case, (\ref{TD_model_vec}) can be rewritten as
\begin{align}
{\bf{r}}_t = {\bf{H}}_{{\rm{T}},{t}} {\bf{s}}_t + {\bf{w}}_t, \label{TD_model_vec_t}
\end{align}
where subscript $t$ denotes the frame index and
\begin{align}
{\bf{H}}_{{\rm{T}},{t}} \buildrel \Delta \over = \sum\limits_{p = 1}^P {{h_{p,t}}{e^{ - j2\pi \frac{{{\tilde{k}_{p,t}}{l_{p,t}}}}{{MN}}}}{{\bm \Delta} ^{{\tilde{k}_{p,t}}}}{{\bf{\Pi }}^{{l_{p,t}}}}} \label{H_T_t}
\end{align}
denotes the effective time domain channel matrix at frame $t$.
Note that~\eqref{TD_model_vec_t} aligns with the time domain input-output relationship given in~\cite{Raviteja2019practical}, and they can be shown to be equivalent in the case of integer delay and Doppler. Then, the received DD domain symbol vector can be obtained based on~\eqref{TD_model_vec_t}, which is given by \cite{li2021cross}
\begin{align}
{{\bf{y}}_{{\rm{DD}},{t}}} &= \left( {{{\bf{F}}_N} \otimes {{\bf{I}}_M}} \right){\bf{r}}_t\notag\\
&= {{\bf{H}}_{{\rm{DD}},{t}}}{{\bf{x}}_{{\rm{DD},{t}}}} + {{\bf{w}}_{{\rm{DD}},{t}}}, \label{DD_model_vec}
\end{align}
where ${{\bf{y}}_{{\rm{DD}},{t}}} = {\rm{vec}}\left( {{{\bf{Y}}_{{\rm{DD}},{t}}}} \right) \in \mathbb{C}^{MN \times 1}$ is the DD domain received symbol vector, ${{\bf{w}}_{{\rm{DD}},{t}}} \sim \mathcal{CN}(\mathbf{0},\sigma^2\mathbf{I}_{MN})$ is the corresponding DD domain CSCG vector with $\sigma^2$ being the noise variance, and ${{\bf{H}}_{{\rm{DD}},{t}}} \in \mathbb{C}^{MN \times MN} $ is the effective DD domain channel matrix given by
\begin{align}
{{\bf{H}}_{{\rm{DD}},{t}}} &= \left( {{{\bf{F}}_N} \otimes {{\bf{I}}_M}} \right){{\bf{H}}_{{\rm{T}},{t}}}\left( {{\bf{F}}_N^{\rm{H}} \otimes {{\bf{I}}_M}} \right)\notag\\
&=\left( {{{\bf{F}}_N} \otimes {{\bf{I}}_M}} \right)\sum\limits_{p = 1}^P {{h_{p,t}}{e^{ - j2\pi \frac{{\tilde{k}_{p,t}{l_{p,t}}}}{{MN}}}}{{\bm \Delta} ^{\tilde{k}_{p,t}}}{{\bf{\Pi }}^{{l_{p,t}}}}} \left( {{\bf{F}}_N^{\rm{H}} \otimes {{\bf{I}}_M}} \right).
\label{H_DD}
\end{align}

\section{Problem Formulation}
In this section, we first develop a transmission protocol for predictive precoder based on the system model in Section II.
According to the proposed transmission protocol, we then formulate the problem of predictive precoder design for the considered OTFS-enabled URLLC system.
The details will be introduced in the following.

\begin{figure}
\centering
\includegraphics[width=0.8\textwidth]{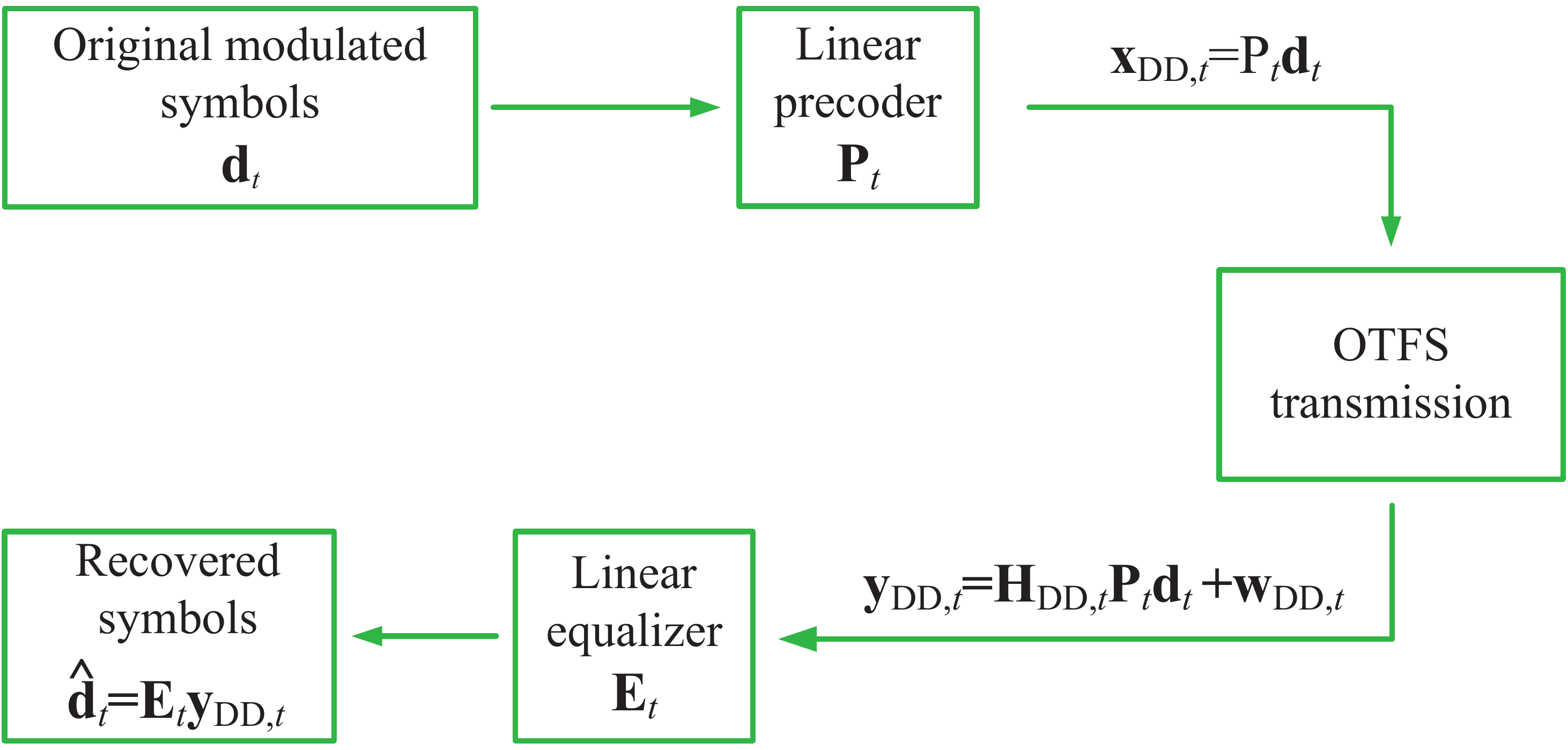}
\caption{The adopted precoder model for OTFS transmission.}
\label{precoder_model}
\centering
\end{figure}

\subsection{Predictive Precoder-based Transmission Protocol}
As depicted in Fig. \ref{precoder_model}, we adopt a precoder at the transmitter before the module of the OTFS transmission to adapt to the DD domain channel matrix to further reduce the FER at the receiver for improving the reliability of the considered OTFS-enabled URLLC system.
Denote by $\mathbf{P}_t \in \mathbb{C}^{MN \times K}$ the precoder matrix at frame $t$, we can exploit $\mathbf{P}_t$ to transform the $K$-length, $K \leq MN$\footnotemark\footnotetext{For ease of study, we call the cases of $K=MN$ and $K<MN$ as the full and dropping modes, respectively, as commonly adopted in e.g., \cite{HuangQin2022ComLet}.
Note that the dropping mode indicates a lower data rate than that of the full mode and we can flexibly adjust the data rate by varying the value of $K$.}, modulated information symbols $\mathbf{d}_t$ to the $MN$-length information symbols $\mathbf{x}_{{\mathrm{DD}},t}$, i.e.,
\begin{equation}\label{x_DD_t}
  \mathbf{x}_{{\mathrm{DD}},t} = \mathbf{P}_t\mathbf{d}_t,
\end{equation}
where the symbols in $\mathbf{d}_t$ are assumed to be independent with unit power, i.e., $\mathbb{E}[\mathbf{d}_t\mathbf{d}_t^H] = \mathbf{I}_{K}$.
In this case, the transmit power can be calculated by $\mathbb{E}[\mathrm{tr}(\mathbf{x}_{{\mathrm{DD}},t}\mathbf{x}_{{\mathrm{DD}},t}^H)] = \mathrm{tr}(\mathbf{P}_t\mathbf{P}_t^H)$.
Then, given a received $MN$-length symbol vector which is expressed as
\begin{equation}\label{}
  \mathbf{y}_{{\mathrm{DD}},t}=\mathbf{H}_{{\mathrm{DD}},{t}}\mathbf{P}_t\mathbf{d}_t + {{\bf{w}}_{{\rm{DD}},{t}}}.
\end{equation}
According to Fig. \ref{precoder_model}, we can adopt a linear equalizer $\mathbf{E}_t \in \mathbb{C}^{MN \times MN}$ to obtain the recovered transmitted symbols $\hat{\mathbf{d}}_t$, which is given by
\begin{equation}\label{d_hat}
  \hat{\mathbf{d}}_t = \mathbf{E}_t \mathbf{y}_{{\mathrm{DD}},t}.
\end{equation}
Generally, there are two classical equalization methods, i.e., ZF and MMSE schemes \cite{tse2005fundamentals, kay1993fundamentals}, and they can be integrated into an unified form which is given by
\begin{equation}\label{E_t}
  \mathbf{E}_t = \left( \varsigma\sigma^2 \mathbf{I}_{MN} + \mathbf{P}_t^H\hat{\mathbf{H}}_{{\mathrm{DD}},{t}}^H\hat{\mathbf{H}}_{{\mathrm{DD}},{t}}\mathbf{P}_t \right)^{-1}\mathbf{P}_t^H\hat{\mathbf{H}}_{{\mathrm{DD}},{t}}^H
\end{equation}
where $\varsigma = 0$ for ZF equalization and $\varsigma = 1$ for MMSE equalization, and $\hat{\mathbf{H}}_{{\mathrm{DD}},{t}}$ denotes the estimated channel matrix.
In this case, the signal-to-interference-plus-noise ratio (SINR) at the receiver with respect to (w.r.t.) the $k$-th symbol in frame $t$, can be calculated as
\begin{equation}\label{SINR}
  \mathrm{SINR}_{k,t} = \frac{|[\mathbf{E}_t\mathbf{H}_{{\mathrm{DD}},{t}}\mathbf{P}_t]_{k,k}|^2}
  {|[\mathbf{E}_t\mathbf{H}_{{\mathrm{DD}},{t}}\mathbf{P}_t]_{k,:}-[\mathbf{E}_t\mathbf{H}_{{\mathrm{DD}},{t}}\mathbf{P}_t]_{k,k}|^2 + \sigma^2|[\mathbf{E}_t]_{k,:}|^2}.
\end{equation}
According to \cite{palomar2005minimum, palomar2003joint}, under a Gray encoding and an $M_{\mathrm{Mod}}$-ary quadrature amplitude modulation (QAM) constellation, the symbol error rate (SER) w.r.t. the $k$-th symbol in frame $t$ can be theoretically derived as a function of SINR, which is given by
\begin{equation}\label{SER}
  \mathrm{SER}_{k,t} \approx \alpha\,\mathrm{erfc} \left(\sqrt{\beta\mathrm{SINR}_{k,t}}\right),
\end{equation}
where $\alpha = \left(2 - 2/\sqrt{M_{\mathrm{Mod}}}\right)/\log_2 M_{\mathrm{Mod}}$, $\mathrm{erfc}(\cdot)$ is the complementary error function, and $\beta = 3/\left(2\sqrt{M_{\mathrm{Mod}}} - 2 \right)$.
Considering that the FER is usually adopted as a metric of evaluating the reliability of the URLLC system, we then derive the associated FER expression w.r.t. frame $t$, i.e.,
\begin{equation}\label{FER}
  \mathrm{FER}_t = 1 - \Pi_{k=1}^K(1 - \mathrm{SER}_{k,t}).
\end{equation}

\textbf{Remark 1}:
Note that for a given channel matrix, the derived theoretical FER expression in (\ref{FER}) matches perfectly with the simulation result \cite{tse2005fundamentals, palomar2005minimum}.
To obtain the theoretical FER expression at a given signal-to-noise ratio (SNR), we can adopt a Monte Carlo method \cite{rubinstein2016simulation} to approximate the simulation result via averaging sufficient large numbers of theoretical FER results calculated based on various channel realizations.
Thus, the derived mathematical FER expression in (\ref{FER}) can be adopted in the considered OTFS-enabled URLLC system for characterizing the system reliability and its efficiency will be verified via simulations in Section VI.

\begin{figure}[t]
  \centering
  \includegraphics[width=\linewidth]{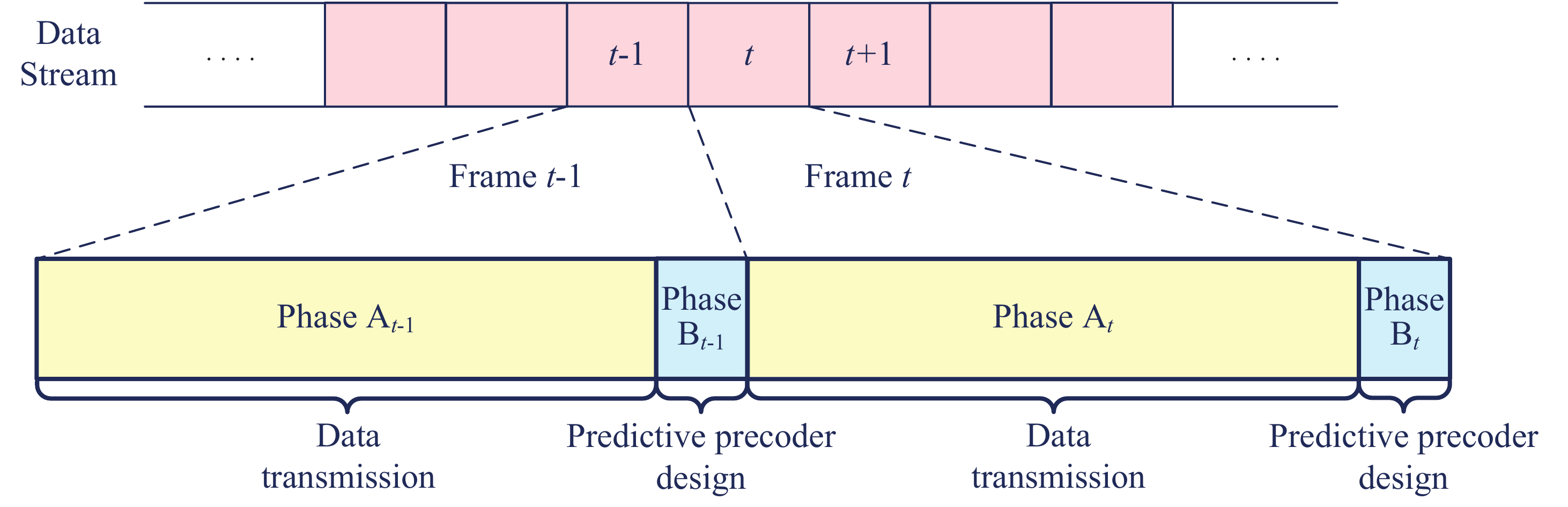}
  \caption{Illustration of the proposed transmission protocol for the considered OTFS-enabled URLLC system.}\label{Fig:protocol_structure}
\end{figure}

According to (\ref{E_t})-(\ref{FER}), we observe that the design of $\mathbf{P}_t$ highly depends on the ICSIT, i.e., $\mathbf{H}_{{\mathrm{DD}},{t}}$ at the transmitter, which however, is not available since the uplink channel estimation at the beginning of each frame is not always available in a rapidly time-varying channel environment.
It is worth noting that if we can exploit the historical channels to predict the required precoder in advance, we can not only bypass the acquisition of the ICSIT, but also guarantee the system FER performance in the subsequent period.
Inspired by this, as shown in Fig. \ref{Fig:protocol_structure}, we develop a predictive precoder-based transmission protocol, where each frame $t$ consists of two phases, i.e., Phase A$_t$: Data transmission and Phase B$_t$: Predictive precoder design.
In Phase A$_t$, the transmitter can directly send the signals to the receiver by exploiting the designed predictive precoder from Phase B$_{t-1}$.
After data transmission, in Phase B$_t$, by leveraging the estimated historical channels\footnotemark\footnotetext{The estimated historical channels can be obtained via the feedback link from the receiver to the transmitter.}, i.e., $\hat{\mathbf{H}}_{{\mathrm{DD}},t-1}, \hat{\mathbf{H}}_{{\mathrm{DD}},t-2}, \cdots, \hat{\mathbf{H}}_{{\mathrm{DD}},t-\tau}$, $\forall t > \tau $, the transmitter can exploit the temporal dependency of these sequential channels to facilitate the predictive precoder design, i.e., predicting the precoder for frame $t+1$ to grant a satisfactory FER performance.


\subsection{Predictive Precoder Problem Formulation}
Based on the developed protocol and Remark 1, in this paper, we aim to minimize the expectation of FER expression of frame $t$, $t \in \mathcal{T} = \{t| t \geq \tau + 1, t \in \mathbb{N}_1 \} $ given the estimated historical channels of past $\tau$ frames via optimizing the predictive precoder at the transmitter subject to the transmission power constraint.
Submitting (\ref{SINR}) and (\ref{SER}) into (\ref{FER}), we can derive the expression of FER in terms of $\mathbf{H}_{\mathrm{DD},{t}}$ and ${\mathbf{P}}_t$, which is given by
\begin{equation}\label{}
  f_{\mathrm{FER}}(\mathbf{H}_{{\mathrm{DD}},{t}},\mathbf{P}_t) \triangleq 1 - \Pi_{k=1}^K\left(1 - \alpha \,  \mathrm{erfc}\left(\sqrt{\frac{\beta|[\mathbf{E}_t\mathbf{H}_{{\mathrm{DD}},{t}}\mathbf{P}_t]_{k,k}|^2}
  {|[\mathbf{E}_t\mathbf{H}_{{\mathrm{DD}},{t}}\mathbf{P}_t]_{k,:}-[\mathbf{E}_t\mathbf{H}_{{\mathrm{DD}},{t}}\mathbf{P}_t]_{k,k}|^2 + \sigma^2|[\mathbf{E}_t]_{k,:}|^2}}\right)\right).
\end{equation}
Then, the associated predictive precoder problem formulation can be expressed as
\begin{align}
(\mathcal{P}):~\mathop{\max}\limits_{{\mathbf{P}}_t } ~ &\mathbb{E}_{\mathbf{H}_{\mathrm{DD},{t}}|\mathcal{H}_{\mathrm{DD},{t}}^{\tau}}
\left\{f_{\mathrm{FER}}(\mathbf{H}_{{\mathrm{DD}},{t}},{\mathbf{P}}_t)
\right\}  \label{PF_objective} \\
\mathrm{s.t.}~&\|\mathbf{P}_t\|_F^2\leq P_0, \forall t \in \mathcal{T} . \label{PF_power}
\end{align}
Here, ${\mathbf{P}}_t \in \mathbb{C}^{MN \times K}$ is the designed precoder for frame $t$, as defined in (\ref{x_DD_t}). $\mathbb{E}_{\mathbf{H}_{\mathrm{DD},{t}}|\mathcal{H}_{\mathrm{DD},{t}}^{\tau}}\{\cdot\}$ denotes the statistical expectation w.r.t. $\mathbf{H}_{\mathrm{DD},{t}}$ given the estimated historical channels $\mathcal{H}_{\mathrm{DD},{t}}^{\tau} \triangleq [\hat{\mathbf{H}}_{\mathrm{DD},{t-1}},\hat{\mathbf{H}}_{\mathrm{DD},{t-2}},\cdots,\hat{\mathbf{H}}_{\mathrm{DD},{t-\tau}}]$.
$P_0$ in constraint (\ref{PF_power}) is the required transmission power budget for an arbitrary frame $t$.
In general, it is almost intractable to directly handle problem $(\mathcal{P})$ since deriving the closed-form of (\ref{PF_objective}) is quite challenging if not impossible.
Furthermore, even if the perfect ICSIT is available, the objective function is non-convex w.r.t. ${\mathbf{P}}_t$ due to the sophisticated expression in $f_{\mathrm{FER}_k}(\mathbf{H}_{{\mathrm{DD}},{t}},{\mathbf{P}}_t)$.
As an alternative, a DL approach will be adopted to exploit the powerful data-driven capability to address this challenging optimization problem in $(\mathcal{P})$ to further improve the system performance.


\begin{figure}[t]
  \centering
  \includegraphics[width=\linewidth]{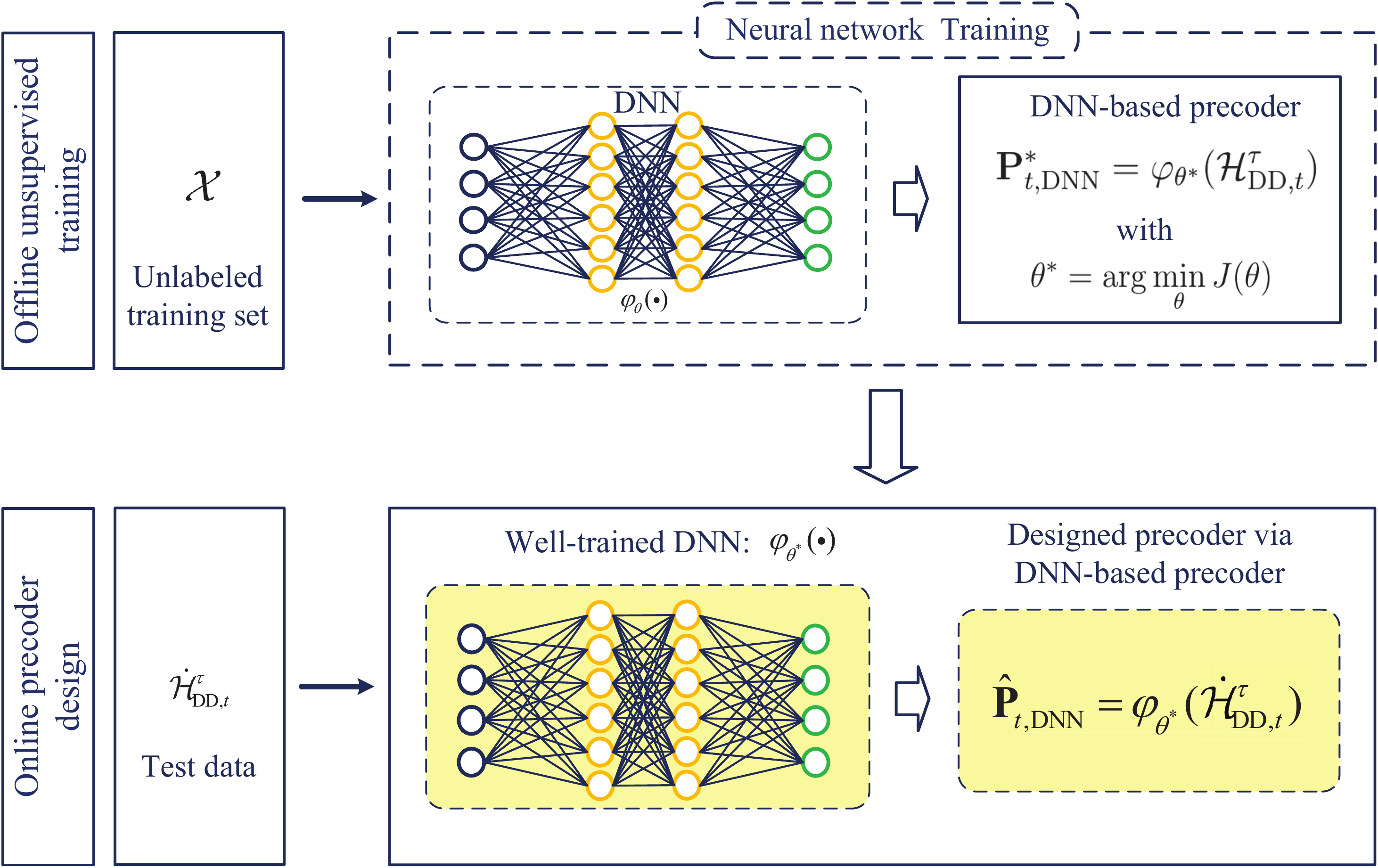}
  \caption{Illustration of the developed DL-based predictive precoder framework for OTFS-enabled URLLC.}\label{Fig:DNN_framework}
\end{figure}

\section{DL-based Predictive Precoder Framework for OTFS-Enabled URLLC}
In this section, we develop a DL-based framework to implicitly learn the features of subsequent channels from the estimated historical channels for the design of predictive precoder.
As illustrated in Fig. \ref{Fig:DNN_framework}, the developed framework consists of offline unsupervised training and online precoder design, as will be introduced in the following.

\subsection{Offline Unsupervised Training}
Let
\begin{equation}\label{}
  \mathcal{X} = \left\{(\mathcal{H}_{{\mathrm{DD}},{t}}^{\tau(1)},\mathbf{H}_{{\mathrm{DD}},{t}}^ {(1)} ), (\mathcal{H}_{{\mathrm{DD}},{t}}^{\tau(2)},\mathbf{H}_{{\mathrm{DD}},{t}}^ {(2)} ), \cdots, (\mathcal{H}_{{\mathrm{DD}},{t}}^{\tau(N_t)},\mathbf{H}_{{\mathrm{DD}},{t}}^ {(N_t)} ) \right\},
\end{equation}
denote the unlabeled training set, where $(\mathcal{H}_{{\mathrm{DD}},{t}}^{\tau(i)},\mathbf{H}_{{\mathrm{DD}},{t}}^ {(i)} )$ is the $i$-th, $i \in \{1,2,\cdots,N_t\}$, training example and $N_t$ denotes the total number of unlabeled training examples of $\mathcal{X}$.
As depicted in Fig. \ref{Fig:DNN_framework}, the training set is then sent to the DNN for neural network training.
Given $\mathcal{H}_{{\mathrm{DD}},{t}}^{\tau(i)}$, the output of the adopted DNN can be expressed as
\begin{equation}\label{}
\mathbf{P}_{t,\mathrm{DNN}}^{(i)} =  \varphi_{\theta}(\mathcal{H}_{{\mathrm{DD}},{t}}^{\tau(i)}),
\end{equation}
where $\varphi_{\theta}(\cdot)$ is the mathematical function expression of the adopted DNN with parameters $\theta$.
Note that although the objective function in (\ref{PF_objective}) is a sophisticated statistical expectation, we can adopt a Monte Carlo method \cite{rubinstein2016simulation} to approximate it, i.e.,
\begin{equation}\label{objective_app}
  \mathbb{E}_{\mathbf{H}_{\mathrm{DD},{t}}|\mathcal{H}_{\mathrm{DD},{t}}^{\tau}}
\left\{f_{\mathrm{FER}}(\mathbf{H}_{{\mathrm{DD}},{t}},{\mathbf{P}}_t)
\right\} \approx \frac{1}{N_t}\sum_{i=1}^{N_t}f_{\mathrm{FER}}\left(\mathbf{H}_{\mathrm{DD},{t}}^{(i)},\mathbf{P}_{t,\mathrm{DNN}}^{(i)}\right).
\end{equation}
Since this is an unsupervised learning problem, the cost function can be directly set as minimizing (\ref{objective_app}) which is given by
\begin{equation}\label{cost_function_framework}
  J(\theta) = \frac{1}{N_t}\sum_{i=1}^{N_t}f_{\mathrm{FER}}\left(\mathbf{H}_{\mathrm{DD},{t}}^{(i)},\varphi_{\theta}(\mathcal{H}_{{\mathrm{DD}},{t}}^{\tau(i)})\right).
\end{equation}
Based on this, we can then operate the unsupervised training via the backpropagation (BP) algorithm to update the neural network parameters progressively, resulting in a well-trained DNN which is expressed as
\begin{equation}\label{DNN_output}
  \mathbf{P}_{t,\mathrm{DNN}^*} =  \varphi_{\theta^*}(\mathcal{H}_{{\mathrm{DD}},{t}}^{\tau}),
\end{equation}
where $\mathbf{P}_{t,\mathrm{DNN}}^*$ denotes the DNN-based precoder based on an arbitrary input $\mathcal{H}_{{\mathrm{DD}},{t}}^{\tau}$, and
\begin{equation}\label{}
  \theta^*=\arg \mathop{\min}\limits_{\theta} J(\theta)
\end{equation}
represents the well-trained neural network parameters.

\begin{figure}[t]
  \centering
  \includegraphics[width=\linewidth]{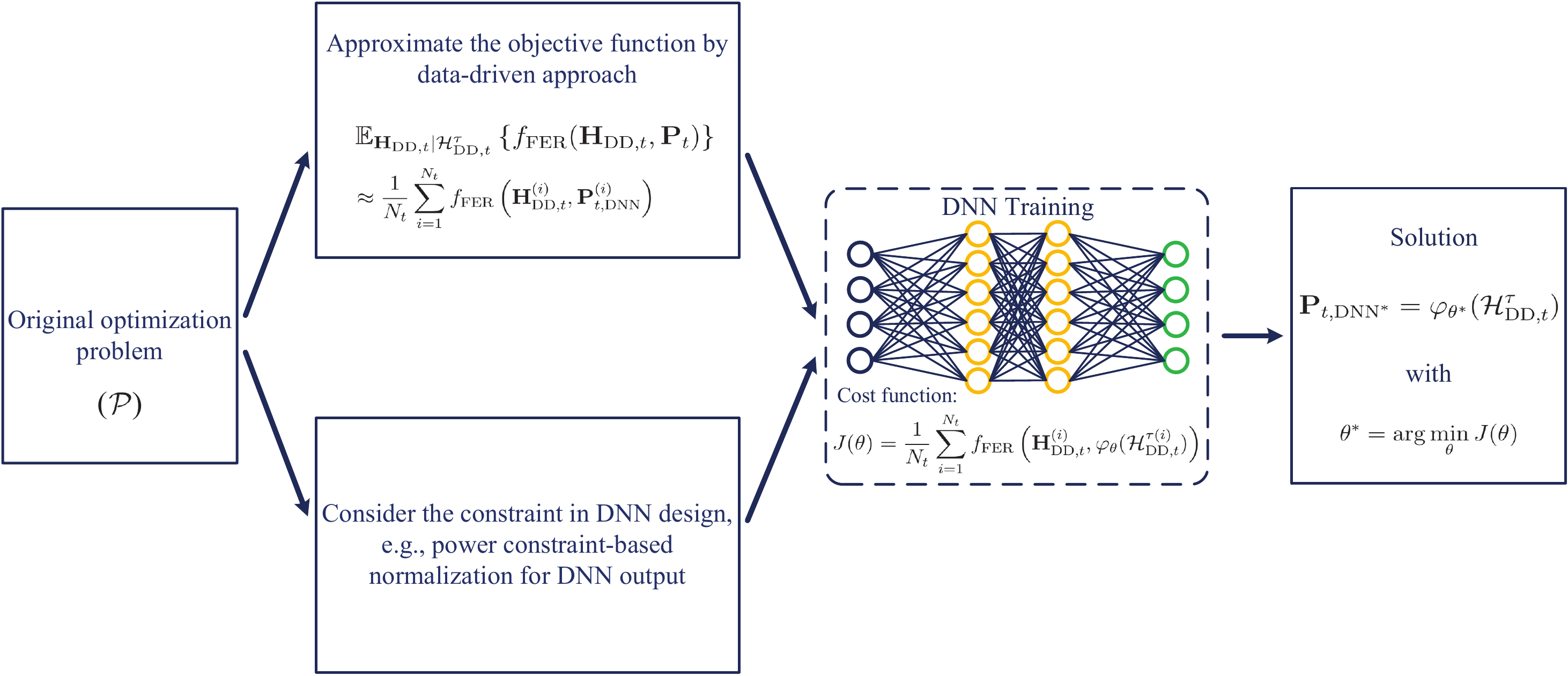}
  \caption{A flowchart of the design of DL-based predictive precoder framework for OTFS-enabled URLLC.}\label{Fig:DL_framework}
\end{figure}

To obtain more insight, we provide a flowchart of the design of DL-based predictive precoder framework in Fig. \ref{Fig:DL_framework}.
Given an original optimization problem ($\mathcal{P}$) as defined in (\ref{PF_objective}), we first transform the objective function and the constraints into appropriate forms to facilitate DNN processing.
For the objective function, we adopt a Monte Carlo approach to approximate it via exploiting the training set.
For the associated power constraint in ($\mathcal{P}$), we adopt the normalization method \cite{goodfellow2016deep, jiang2021learning} in the designed DNN to generate the desired output, which will be detailed in the next section.
Based on this, we can then design a DNN with appropriate neural network structure and cost function.
Then, through DNN training, we can finally obtain the desired solution, i.e., the predictive precoder as defined in (\ref{DNN_output}).

\subsection{Online Precoder Design}
Based on the well-trained neural network obtained from the offline training, we can then execute the online precoder design.
Given a test example, denoted by $\dot{\mathcal{H}}_{{\mathrm{DD}},{t}}^{\tau}$, we can directly send it to $\varphi_{\theta^*}(\cdot)$ to obtain the designed precoder, which is given by
\begin{equation}\label{}
  \hat{\mathbf{P}}_{t,\mathrm{DNN}^*} =  \varphi_{\theta^*}(\dot{\mathcal{H}}_{{\mathrm{DD}},{t}}^{\tau}).
\end{equation}

\textbf{Remark 2}:
It is worth noting that the adopted DNN in the developed framework can be realized by any kind of neural network structures, e.g., CNNs \cite{liu2019deep}, recurrent neural networks \cite{goodfellow2016deep, hochreiter1997long}, residual neural networks \cite{liu2022deepresidual}, etc.
Thus, the proposed framework in Fig. \ref{Fig:DNN_framework} is a universal predictive precoder design framework for OTFS-enabled URLLC.
By exploiting both the predictive mechanism and the powerful data-driven capability of DL, the developed framework can further improve the system reliability without the availability of ICSIT.

\section{DDC-aware CLSTM Neural Network for Predictive Precoder Design}
This section provides a realization of the developed framework.
In particular, we exploit the powerful CLSTM neural network to extract the spatial-temporal features from estimated historical channels to serve the predictive precoder design task, resulting a novel DDCL-Net, which facilitates the application of the developed framework.
In the following, we will introduce the proposed DDCL-Net in terms of neural network architecture and associated algorithm, respectively.

\subsection{DDCL-Net Architecture}
\begin{figure}[t]
  \centering
  \includegraphics[width=\linewidth]{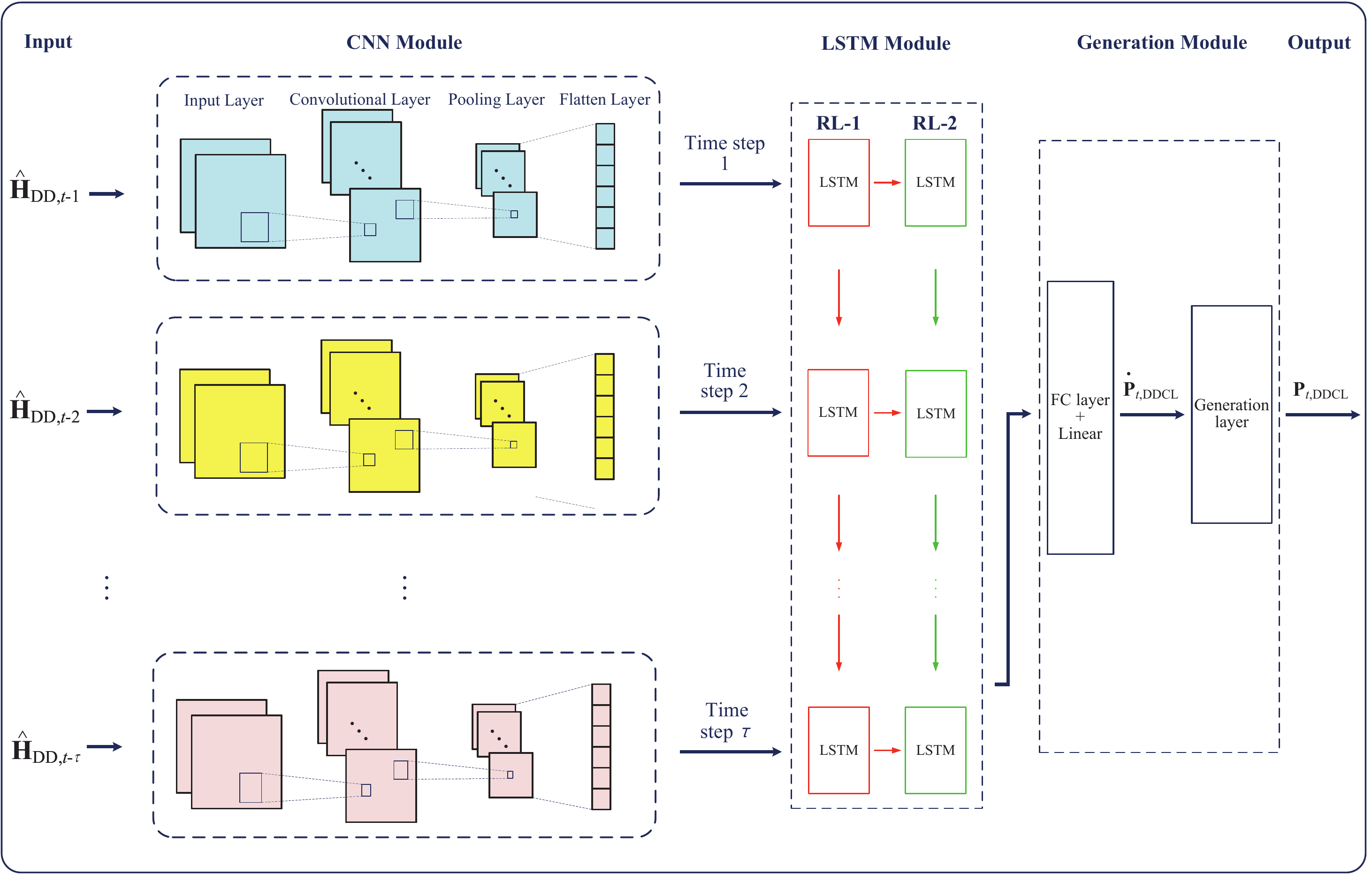}
  \caption{The proposed DDCL-Net architecture for predictive precoder design in the considered OTFS-enabled URLLC system.}\label{Fig:DDCL}
\end{figure}

As depicted in Fig. \ref{Fig:DDCL}, the proposed DDCL-Net consists of five parts: Input, CNN module, LSTM module, generation module, and output, which will be introduced in the following and the associated neural network hyperparameters are presented in Table \ref{Tab:Hyperparameters DDCL-Net}.

\begin{table}[t]
\normalsize
\caption{Neural Network Hyperparameters of the developed DDCL-Net}\vspace{-0.2cm}\label{Tab:Hyperparameters DDCL-Net}
\centering
\small
\renewcommand{\arraystretch}{1.05}
\begin{tabular}{c c c}
  \hline
   \multicolumn{3}{l}{\textbf{Input}: $\tilde{\mathcal{H}}_{\mathrm{DD},{t}}^{\tau} \in \mathbb{R}^{\tau \times MN \times MN \times 2}$}  \\
  \hline
  \multicolumn{3}{l}{\textbf{CNN Module}:} \\
  \hspace{0.1cm} \textbf{Layers} & \textbf{Parameters} &  \hspace{0.3cm} \textbf{Values}   \\
  \hspace{0.1cm} Convolutional layer with ReLU & Size of filters & \hspace{0.3cm}  $ 2 \times 3 \times 3 \times 2$   \\
  \hspace{0.1cm} Pooling layer with maximum pooling & Size of filters & \hspace{0.3cm}  $ 2 \times 2 $   \\
  \hspace{0.1cm} Flatten layer & Output shape & \hspace{0.3cm}  $ 1024 \times 1 $   \\
  \multicolumn{3}{l}{\textbf{LSTM Module}:} \\
  \hspace{0.1cm} \textbf{Layers} & \textbf{Parameters} &  \hspace{0.3cm} \textbf{Values}   \\
  \hspace{0.1cm} RL-1 & Output shape & \hspace{0.3cm}  $ 32 \times \tau $   \\
  \hspace{0.1cm} RL-2 & Output shape & \hspace{0.3cm}  $ 32 \times 1 $   \\
  \multicolumn{3}{l}{\textbf{Generation Module}:} \\
  \hspace{0.1cm} \textbf{Layers} & \textbf{Parameters} &  \hspace{0.3cm} \textbf{Values}   \\
  \hspace{0.1cm}  FC layer & Activation function & \hspace{0.3cm} Linear \\
  \hspace{0.1cm}  FC layer & Output shape & \hspace{0.3cm} $ 2KMN $\\
  \hspace{0.1cm}  Generation layer & Normalization function & \hspace{0.3cm} Power normalization \\
  \hline
   \multicolumn{3}{l}{\textbf{Output}: $\mathbf{P}_{t,\mathrm{DDCL}} \in \mathbb{C}^{MN\times K}$}  \\
  \hline
\end{tabular}
\end{table}

\subsubsection{Input}
According to the framework in Fig. \ref{Fig:DNN_framework}, we adopt the historical channels from frame $t-1$ to frame $t-\tau$ as the input.
Since different historical channel matrices carry  different frame information, they are sequentially handled at different time steps of the DDCL-Net, as illustrated in Fig. \ref{Fig:DDCL}.

\subsubsection{CNN Module}
Inspired by the CNN's excellent space invariant property, we adopt a CNN module to exploit the spatial features of channels from the input to improve the learning performance.
In particular, the CNN module is composed of one input layer, one convolutional layer, one pooling payer, and one flatten layer.
Since the real and imaginary parts of historical channels are independent, we adopt two independent neural network channels to individually handle the real and imaginary parts as the input, i.e.,
\begin{equation}\label{input}
  \tilde{\mathcal{H}}_{\mathrm{DD},{t}}^{\tau} = \mathcal{M}([ \mathrm{Re}\{\mathcal{H}_{\mathrm{DD},{t}}^{\tau}\}, \mathrm{Im}\{\mathcal{H}_{\mathrm{DD},{t}}^{\tau}\} ]),
\end{equation}
where $\tilde{\mathcal{H}}_{\mathrm{DD},{t}}^{\tau}$ is the neural network input and  $\mathcal{M}(\cdot)\!:\mathbb{R}^{MN \times \tau MN}\mapsto\mathbb{R}^{\tau \times MN \times MN \times 2}$ denotes the adopted mapping function.
Based on the historical channels from input layer, the convolutional layer adopts 2 filters to operate convolution to generate the feature maps, where each filter size is set as $3 \times 3 \times 2$ and each convolution is added with a rectified linear unit (ReLU) activation function.
Then, to further reduce the sizes of the generated feature maps, a maximum pooling operation with $2 \times 2$-filter size is adopted in pooling layer.
Finally, a flatten layer is connected to reshape the output of the CNN module for the ease of subsequent processing.

\subsubsection{LSTM Module}
Inspired by LSTM's powerful prediction capability, we adopt an LSTM module in DDCL-Net to exploit the temporal features of historical channels to facilitate the dedicated predictive task.
As depicted in Fig. \ref{Fig:DDCL}, we adopt two recurrent layers (RLs): RL-1 and RL-2 to fully exploit the extracted features from CNN module.
In particular, for each RL, the identical LSTM unit is recurrently adopted at all the $\tau$ time steps to handle the input features from the $\tau$ historical frames.
Furthermore, in RL-1, for each time step, the output of LSTM unit is not only adopted as the input of LSTM unit at the next time step, but also sent to RL-2.
In contrast, in RL-2, the output of LSTM unit generated at each time step is only adopted as the input of LSTM unit at the next time step.
Since RL-2 captures all the temporal dependencies of historical channels, the output of LSTM unit of RL-2 at the last time step, i.e., time step $\tau$, is adopted as the entire output of the LSTM model.

\subsubsection{Generation Module}
The generation module is designed to leverage the extracted features from previous modules to generate the desired output.
As illustrated in Fig. \ref{Fig:DDCL}, the generation module consists of a fully-connected (FC) layer and a generation layer.
In FC layer, we adopt an FC operation with a linear activation function to integrate the extracted features into an appropriate size.
Then, a generation layer is added to refine the FC layer output to meet the power constraint defined in (\ref{PF_power}).
Denote by $\dot{\mathbf{P}}_{t,\mathrm{DDCL}} \in \mathbb{R}^{2MN\times K} $ the real-valued output of the FC layer, where $\dot{\mathbf{P}}_{t,\mathrm{DDCL}}(1:MN,:)$ and $\dot{\mathbf{P}}_{t,\mathrm{DDCL}}(MN+1:2MN,:)$ denote the real and imaginary parts of the associated complex-valued precoder matrix.
In this case, the normalization in terms of power constraint can be expressed as
\begin{equation}\label{P_t_normalization}
  \ddot{\mathbf{P}}_{t,\mathrm{DDCL}} = \sqrt{P_0}\frac{\dot{\mathbf{P}}_{t,\mathrm{DDCL}}}{\|\dot{\mathbf{P}}_{t,\mathrm{DDCL}}\|_F}
  \in \mathbb{C}^{2MN\times 1}.
\end{equation}
Finally, the output of the generation module can be expressed as
\begin{equation}\label{P_t_generation}
  \mathbf{P}_{t,\mathrm{DDCL}} = [\ddot{\mathbf{P}}_{t,\mathrm{DDCL}}(1:MN,:) + j\ddot{\mathbf{P}}_{t,\mathrm{DDCL}}(MN+1:2MN,:)]
  \in \mathbb{C}^{MN\times K},
\end{equation}
where $j = \sqrt{-1}$ denotes the imaginary unit.

\subsubsection{Output}
According to above discussions, the expression of the proposed DDCL-Net can be formulated as
\begin{equation}\label{}
  \mathbf{P}_{t,\mathrm{DDCL}} = g_\omega (\mathcal{H}_{{\mathrm{DD}},{t}}^{\tau}),
\end{equation}
where $g_\omega(\cdot)$ denotes the DDCL-Net with neural network parameters $\omega$.

\textbf{Remark 3}: It is worth noting that the designed DDCL-Net is a universal neural network architecture for precoder design in OTFS-enabled URLLC, and Fig. \ref{Fig:DDCL} and Table \ref{Tab:Hyperparameters DDCL-Net} only provide one of the possible realizations of it.
Through exploiting the powerful scalability of neural networks, our designed DDCL-Net can be easily extended, e.g., by changing the input/output size and number of layers/modules, to fulfil the requirements of various system settings.

\begin{figure}[t]
  \centering
  \includegraphics[width=\linewidth]{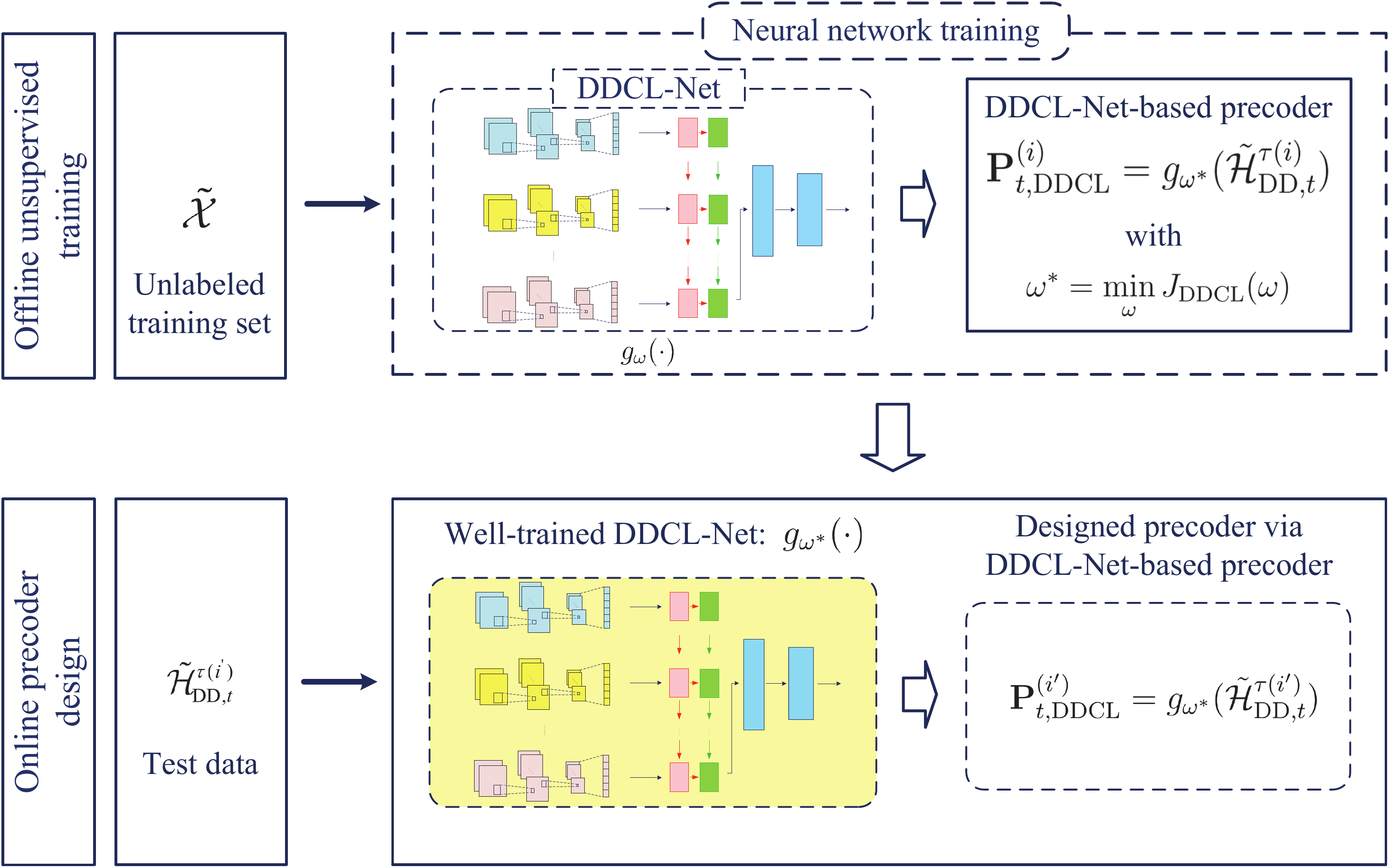}
  \caption{The flowchart of the proposed DDCL-Net-based predictive precoder design algorithm for OTFS-enabled URLLC.}\label{Fig:DDCL_algorithm}
\end{figure}

\subsection{DDCL-Net-based Predictive Precoder Design Algorithm}
In this section, we apply the designed DDCL-Net to the developed framework to propose a novel DDCL-Net-based predictive precoder design algorithm, as illustrated in Fig. \ref{Fig:DDCL_algorithm}.
The proposed algorithm consists of the offline unsupervised training and online precoder design, which will be detailed in the following.

\subsubsection{Offline Unsupervised Training}
Given an unsupervised training set
\begin{equation}\label{}
  \tilde{\mathcal{X}} = \left\{(\tilde{\mathcal{H}}_{{\mathrm{DD}},{t}}^{\tau(1)},\mathbf{H}_{{\mathrm{DD}},{t}}^ {(1)} ), (\tilde{\mathcal{H}}_{{\mathrm{DD}},{t}}^{\tau(2)},\mathbf{H}_{{\mathrm{DD}},{t}}^ {(2)} ), \cdots, (\tilde{\mathcal{H}}_{{\mathrm{DD}},{t}}^{\tau(N_t)},\mathbf{H}_{{\mathrm{DD}},{t}}^ {(N_t)} ) \right\},
\end{equation}
where $\tilde{\mathcal{H}}_{{\mathrm{DD}},{t}}^{\tau(i)}$ is defined according to (\ref{input}),  $(\tilde{\mathcal{H}}_{{\mathrm{DD}},{t}}^{\tau(i)},\mathbf{H}_{{\mathrm{DD}},{t}}^ {(i)} )$ is the $i$-th, $i \in \{1,2,\cdots,N_t\}$, training example and $N_t$ denotes the total number of unlabeled training examples of $\tilde{\mathcal{X}}$.
Thus, based on the developed framework and (\ref{cost_function_framework}), we can derive the cost function of DDCL-Net as
\begin{equation}\label{cost_function_DDCL}
  J_{\mathrm{DDCL}}(\omega) = \frac{1}{N_t}\sum_{i=1}^{N_t}f_{\mathrm{FER}}\left(\mathbf{H}_{\mathrm{DD},{t}}^{(i)},g_\omega(\tilde{\mathcal{H}}_{{\mathrm{DD}},{t}}^{\tau(i)})\right).
\end{equation}
Based on the cost function, we can then leverage the BP algorithm to update the neural network parameters and finally obtain a well-trained DDCL-Net:
\begin{equation}\label{well_trained_DDCL}
  \mathbf{P}_{t,\mathrm{DDCL}}^{(i)} = g_{\omega^*} (\tilde{\mathcal{H}}_{{\mathrm{DD}},{t}}^{\tau(i)}), \forall i,
\end{equation}
where
\begin{equation}\label{}
  \omega^* = \mathop{\min}\limits_{\omega} J_{\mathrm{DDCL}}(\omega)
\end{equation}
denotes the well-trained neural network parameters obtained from offline training and $g_{\omega^*}(\cdot)$ denotes the well-trained DDCL-Net with $\omega^*$.

\begin{table}[t]
\small
\centering
\begin{tabular}{l}
\toprule[1.8pt] \vspace{-0.6 cm}\\
\hspace{-0.1cm} \textbf{Algorithm 1} {DDCL-Net-based Predictive Precoder Design Algorithm} \vspace{0.2 cm} \\
\toprule[1.8pt] \vspace{-0.6 cm}\\
\textbf{Parameters Initialization:} $i_t = 0$ \\
\hspace{4.0cm}Random neural network parameters $\omega$ \\
\hspace{4.0cm}Offline training set $\tilde{\mathcal{X}}$ \\
\textbf{Offline Unsupervised Training:} \\
1:\hspace{0.75cm}\textbf{Input:} Training set $\tilde{\mathcal{X}}$\\
2:\hspace{1.1cm}\textbf{while} $i_t \leq N_{\mathrm{tr}} $ \textbf{do} \\
3:\hspace{1.6cm}Progressively update $\varsigma$ to minimize $J_{\mathrm{DDCL}}(\omega)$ in (\ref{cost_function_DDCL}) based on BP algorithm  \\
\hspace{1.8cm} $i_t = i_t + 1$  \\
4:\hspace{1.1cm}\textbf{end while} \\
5:\hspace{0.75cm}\textbf{Output}:  Well-trained $g_{\omega^*}( \cdot ) $ as defined in (\ref{well_trained_DDCL})\\
\textbf{Online Precoder Design:} \\
6:\hspace{0.75cm}\textbf{Input:} Test data  $\tilde{\mathcal{H}}_{{\mathrm{DD}},{t}}^{\tau(i')}$ as defined in (\ref{test_precoder})  \\
7:\hspace{1.1cm} Calculate predictive precoder based on $g_{\omega^*}(\cdot)$ \\
8:\hspace{0.75cm}\textbf{Output:} $\mathbf{P}_{t,\mathrm{DDCL}}^{(i')} = g_{\omega^*} (\tilde{\mathcal{H}}_{{\mathrm{DD}},{t}}^{\tau(i')})$ \vspace{0.2cm}\\
\bottomrule[1.8pt]
\end{tabular}
\end{table}

\subsubsection{Online Precoder Design}
As shown in Fig. \ref{Fig:DDCL_algorithm}, after obtaining the well-trained DDCL-Net, in online phase, we can directly send test examples for predictive precoder design, i.e.,
\begin{equation}\label{test_precoder}
  \mathbf{P}_{t,\mathrm{DDCL}}^{(i')} = g_{\omega^*} (\tilde{\mathcal{H}}_{{\mathrm{DD}},{t}}^{\tau(i')}),
\end{equation}
where $\tilde{\mathcal{H}}_{{\mathrm{DD}},{t}}^{\tau(i')}$ denotes an arbitrary test example with $i' \neq i$.

\subsubsection{Algorithm Steps}
According to the above discussion, we summarize the predictive precoder design algorithm steps in Algorithm 1, where $N_{\mathrm{tr}}$ denotes the total number of iterations in offline training and can be set according to the early stopping criteria \cite{goodfellow2016deep}.

\begin{table}[t]
\normalsize
\caption{Simulation Settings}\label{Tab:Simulation_settings}
\vspace{-0.3cm}
\centering
\small
\renewcommand{\arraystretch}{1.05}
\begin{tabular}{c c}
  \hline
  \vspace{-0.6cm} \\
   \textbf{Parameters} & \hspace{0.4cm} \textbf{Default Values} \hspace{0.4cm} \\
  \hline
  \vspace{-0.6cm} \\
  \vspace{-0.6cm} \\
   The number of delay bins/number of subcarriers & $M = 8$  \\
  \vspace{-0.6cm} \\
  \vspace{-0.6cm} \\
   The number of Doppler bins/number of time slots & $N = 4$  \\
  \vspace{-0.6cm} \\
  \vspace{-0.6cm} \\
   Digital modulation scheme & QPSK \\
  \vspace{-0.6cm} \\
  \vspace{-0.6cm} \\
   Carrier frequency & 4 GHz \\
  \vspace{-0.6cm} \\
  \vspace{-0.6cm} \\
   Sub-carrier spacing & 15 kHz \\
  \vspace{-0.6cm} \\
  \vspace{-0.6cm} \\
   Maximum delay & $l_{\max}=5$ \\
  \vspace{-0.6cm} \\
  \vspace{-0.6cm} \\
   Maximum Doppler-shift & $k_{\max}=2$ \\
  \vspace{-0.6cm} \\
  \vspace{-0.6cm} \\
  Number of resolvable paths & $P = 4 $ \\
  \vspace{-0.6cm} \\
  \hline
\end{tabular}\vspace{-0.6cm}
\end{table}

\section{Numerical Results}
This section provides extensive simulation results for the considered OTFS-enabled URLLC system to verify the effectiveness and efficiency of the developed method.
The default setting of the system are in the following.
As depicted in Fig. \ref{scenario}, there is a single-antenna transmitter serving one single-antenna receiver via OTFS transmission with $M = 8$, $N = 4$.
Without loss of generality, the system adopts a quadrature phase shift keying (QPSK) signaling with a carrier frequency of 4 GHz, and the sub-carrier is spaced by a bandwidth of 15 kHz.
In addition, the wireless channel is generated based on (\ref{H_DD}), and the associated delay indices are randomly generated by uniform distribution on $[0,l_{\max}]$ and $[-k_{\max},k_{\max}]$, respectively.
In particular, we consider an LTV channel with $P=4$ resolvable paths and $l_{\max}=5$, and the mobile receiver is with a maximum speed of $100~\mathrm{km/h}$, resulting in a $k_{\max} = 2$.
The fading coefficients are independently generated by a complex Gaussian distribution with a zero-mean and a variance of $\frac{1}{P}$.
For the formulated predictive precoder problem, we set $\tau = 5$, $v = 1$, and the estimated historical channels for precoder design and the real-time estimated channels for equalization are set with a normalized mean squared error (NMSE) of $0.01$.
For ease of presentation, the default simulation settings are summarized in Table \ref{Tab:Simulation_settings}.

Furthermore, three benchmark schemes are adopted to compare with the developed method to evaluate the performance:
\begin{itemize}
  \item Lower bound: In this scheme, perfect ICSI at both the transmitter and receiver is available and a DL approach is adopted to design the precoder, where a classical CNN is adopted with one convolutional layer, one pooling layer, one flatten layer, and one dense layer.
  By exploiting the perfect ICSI, this DL scheme can achieve an FER performance lower bound of the formulated predictive precoder problem ($\mathcal{P}$).
  \item MMSE scheme: In this method, an MMSE equalizer \cite{tse2005fundamentals, kay1993fundamentals} is adopted at the receiver without taking into the precoder at the transmitter, as defined in equation (\ref{d_hat}) with $\mathbf{P}_t = \mathbf{I}_{K}$, $K = MN$, $\varsigma = 1$.
  \item ZF scheme: In this method, a ZF equalizer \cite{tse2005fundamentals, kay1993fundamentals} is adopted at the receiver without taking into the precoder at the transmitter, as defined in equation (\ref{d_hat}) with $\mathbf{P}_t = \mathbf{I}_{K}$, $K = MN$, $\varsigma = 0$.
\end{itemize}

For the developed DDCL-Net method, the neural network parameters are set according to Table \ref{Tab:Hyperparameters DDCL-Net}, $N_t = 30,000$, $N_{\mathrm{tr}}$ is set based on the Algorithm 1.
For all the considered schemes, unless otherwise specified, all the results are generated through averaging at least $10^6$ Monte Carlo realizations \cite{liu2014maximum}.
The simulation results are provided in terms of three aspects: FER performance, generalization, and reliability-latency tradeoff, which will be introduced in the following.

\begin{figure}[t]
  \centering
  \includegraphics[width=3.6in,height=3.2in]{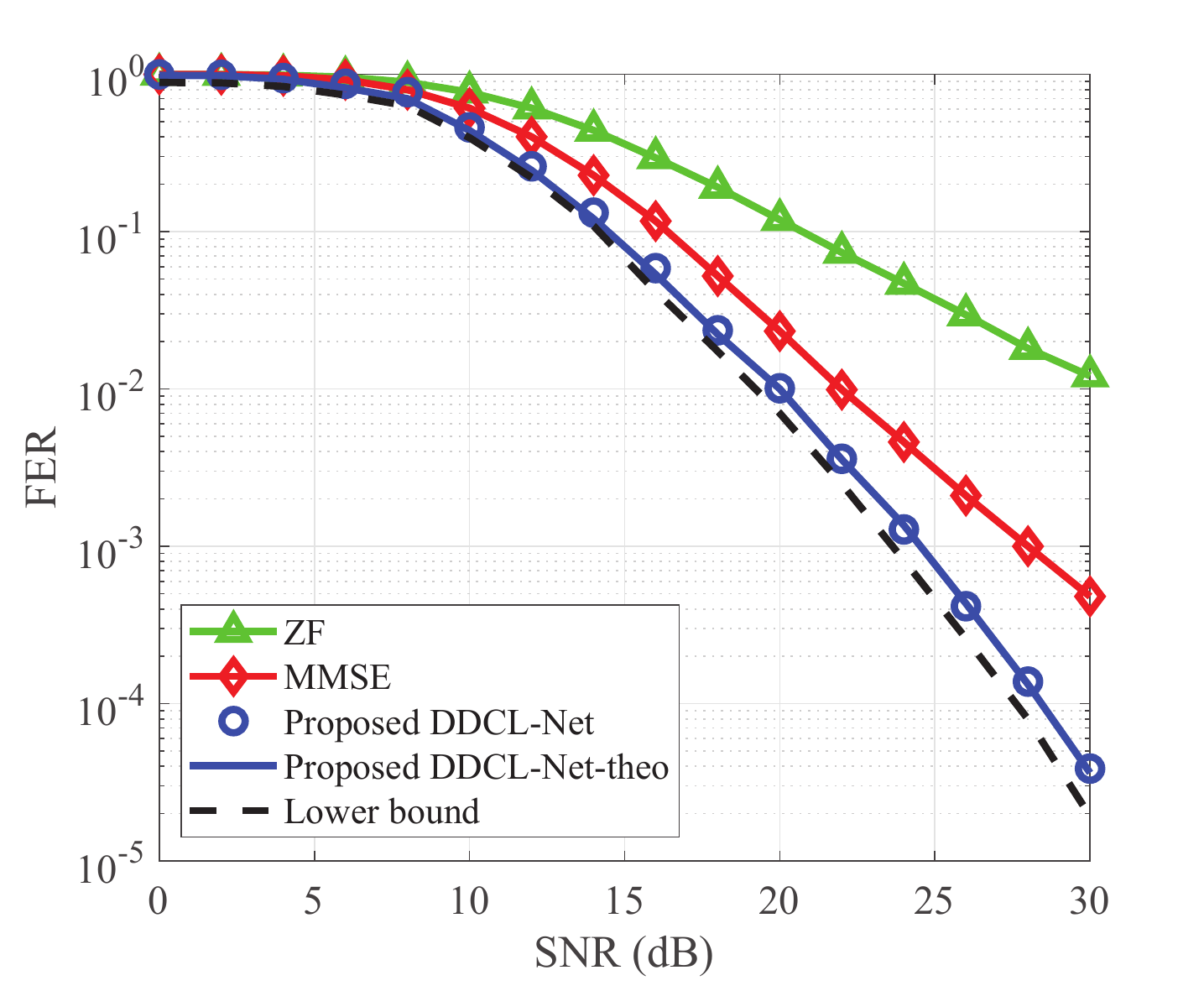}
  \caption{The FER performance with different receive SNRs under $M=8$, $N=4$, $K=32$.}\label{Fig:FER_SNR}
\end{figure}

\subsection{FER Performance}
We first investigate the reliability of the OTFS-enabled URLLC system.
In particular, FER is an important metric to characterize the system reliability and thus we study the FER under different receive SNRs with full model ($K=MN$) and dropping mode ($K<MN$) as defined in (\ref{x_DD_t}).
Fig. \ref{Fig:FER_SNR} presents the curves of FER versus SNR at the receiver under the full mode with QPSK modulation.
On top of the three benchmark methods and the proposed method, the derived theoretical FER expression in (\ref{FER}), denoted by ``DDCL-Net-theo'', is also provided to evaluate its matching performance.
It can be observed that the curves of the proposed DDCL-Net and the DDCL-Net-theo perfectly coincide with each one, which verifies the accuracy of the theoretical FER expression.
We can also find that the FERs of all the considered algorithms decrease with the increase of SNR. This is because a higher SNR indicates a strong signal power in the received samples.
In addition, although the MMSE method outperforms the ZF method, both these two methods only have limited FER performance since they only leverage the equalization for detection without a proper precoder design.
In contrast, the DL-based methods, i.e., the proposed DDCL-Net and lower bound schemes, can achieve a better performance than ZF and MMSE methods.
As depicted in Fig. \ref{Fig:FER_SNR}, the proposed method can achieve a 5 dB SNR gain w.r.t. the $\mathrm{FER}=10^{-4}$ compared with the MMSE method.
In particular, the FER curve of the proposed method decreases in a same slope with that of the lower bound scheme.
These results are as expected since our proposed method can implicitly learn features of the subsequent channels from historical channels for intelligently predicting precoder to guarantee the FER performance in the subsequent period.

\begin{figure}[t]
  \centering
  \includegraphics[width=3.6in,height=3.2in]{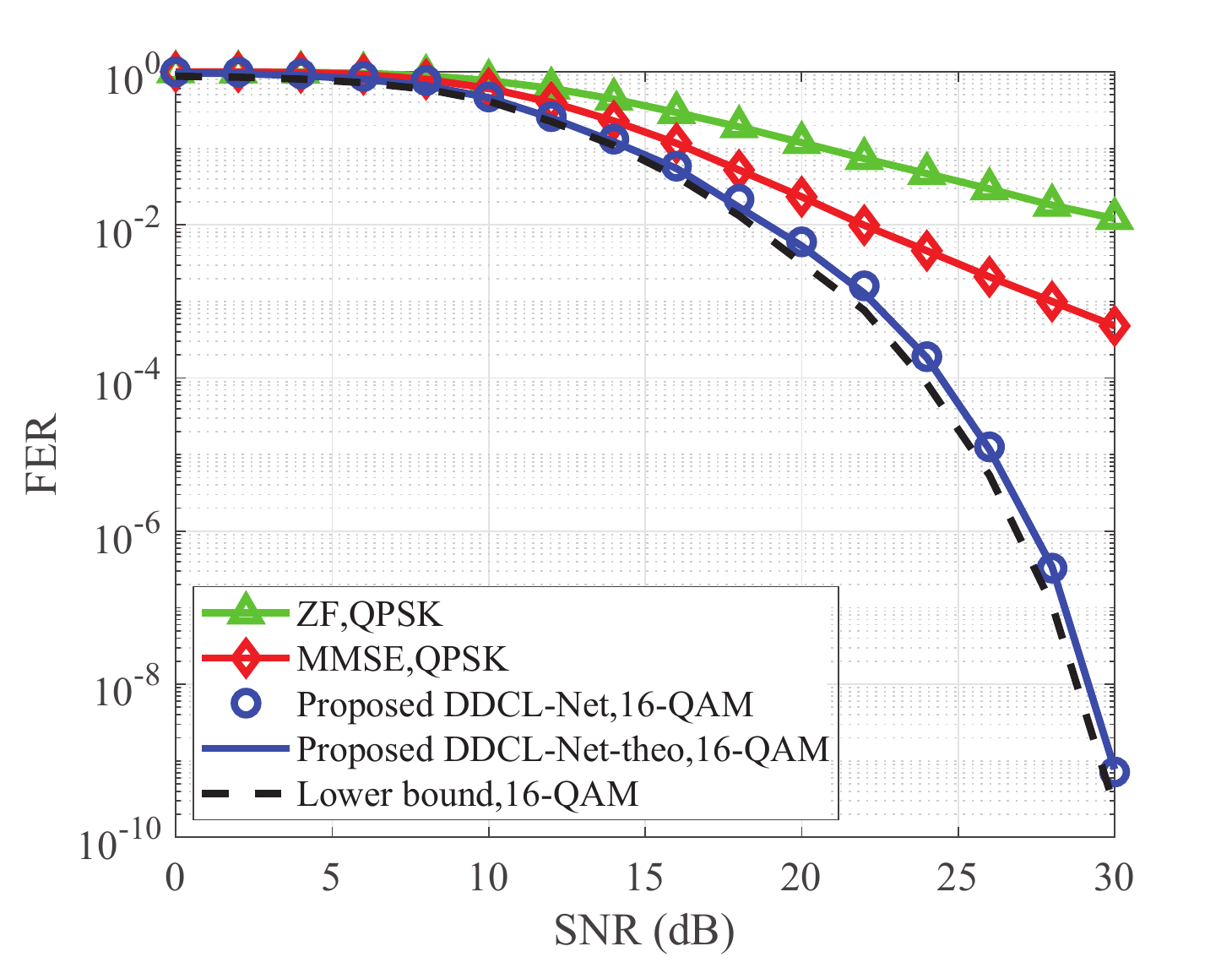}\vspace{-0.2cm}
  \caption{The FER performance with different receive SNRs under $M=8$, $N=4$, $K=16$.}\label{Fig:FER_SNR_QAM}\vspace{-0.2cm}
\end{figure}

Furthermore, we investigate the impact of the dropping mode on FER performance.
Fig. \ref{Fig:FER_SNR_QAM} presents the FER-SNR curve with $K=\frac{1}{2}MN$.
To ensure fair comparisons, the proposed method adopts a 16-QAM modulation scheme to match with $K=\frac{1}{2}MN$, thus guaranteeing the same data rate with that of the QPSK scheme.
It can be observed that the proposed method significantly outperforms the ZF and MMSE methods, and its FER performance is comparable to that of the lower bound method.
Specifically, the proposed method can achieve an FER of less than $10^{-9}$, i.e., a reliability of up to $99.9999999\%$ at an SNR of $30~\mathrm{dB}$, which is $1/100$ times of the required FER level in xURLLC (i.e., $10^{-7}$ \cite{sutton2019enabling}).
This is as expected since the precoder transforms a $\frac{1}{2}MN$-length source symbols to an $MN$-length symbols for transmission, this precoder operation is equivalent to adding some channel coding on the source symbols, which can significantly reduce the system FER.
Meanwhile, our proposed method can exploit DDCL-Net to accurately design the predictive precoder based on the historical channels to further improve the FER performance.

\begin{figure}[t]
  \centering
  \includegraphics[width=3.6in,height=3.2in]{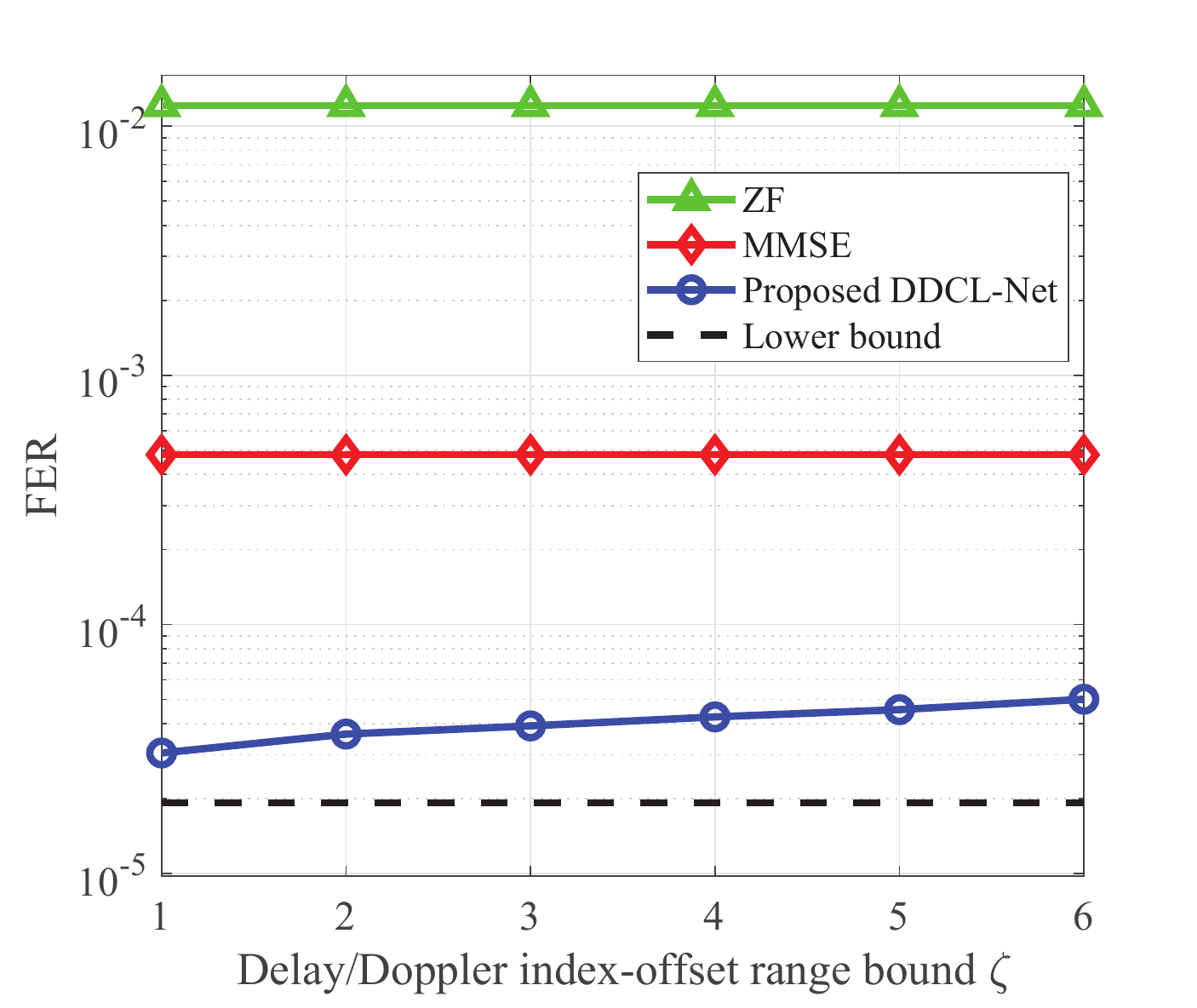}
  \caption{FER versus delay/Doppler index-offset range bound under $K=32$, QPSK modulation, and $\mathrm{SNR = 30~dB}$.}\label{Fig:FER_zeta}
\end{figure}

\begin{figure}[t]
  \centering
  \includegraphics[width=3.6in,height=3.2in]{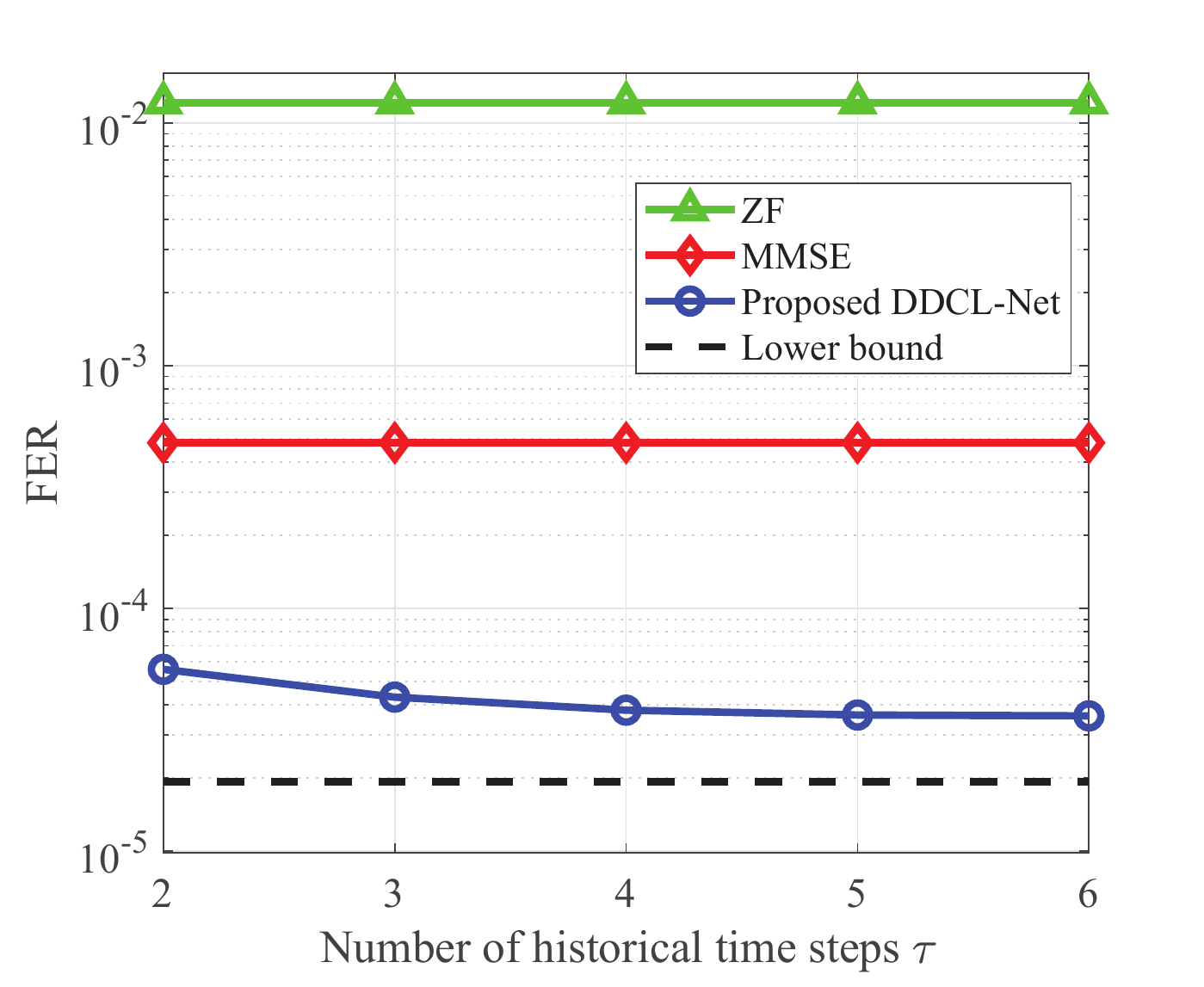}
  \caption{FER versus number of available historical time steps under $K=32$, QPSK modulation, and $\mathrm{SNR = 30~dB}$.}\label{Fig:FER_tau}
\end{figure}

\subsection{Generalizability}
In this section, we investigate the generalizability of the proposed method in terms of the delay index-offset range, Doppler index-offset range, and number of available historical time steps.
For ease of study, we let $\zeta=-\epsilon_{\min}=-\varepsilon_{\min}=\epsilon_{\max}=\varepsilon_{\max}$ denote the offset range bound to generate the associated delay/Doppler index-offset range $[-\zeta,\zeta]$, and the curves of FER versus $\zeta$ is presented in Fig. \ref{Fig:FER_zeta}, where we provide the FER results by varying values of $\zeta \in \{1,2,3,4,5\}$ to characterize different mobility scenarios.
We can find that the FER of the proposed method slightly increases with $\zeta$.
This is inevitable since the large delay/Doppler offset leads to a large difference between the historical CSI and ICSI.
As a remedy, our proposed method can exploit the historical channels to design the predictive precoder to reduce the performance drop when $\zeta$ increases.
In addition, both the ZF, MMSE, and lower bound methods keep constant with different $\zeta$ since these methods do not need to exploit the temporal dependencies for precoder design.

Next, we turn our study to the generalizability in terms of number of available historical time steps, i.e., $\tau$ defined in (\ref{PF_objective}).
Fig. \ref{Fig:FER_tau} presents the FER curves under different numbers of available historical time steps by varying the values of $\tau \in \{2,3,4,5,6\}$.
It is shown that due to the fact that the methods of ZF, MMSE, and lower bound do not leverage the historical channels for precoder design, the FERs of these methods remains constant.
Different from the other schemes, our proposed method is a predictive approach, when $\tau$ increases, it can exploit more historical channels for precoder design to further improve the FER performance, as indicated in Fig. \ref{Fig:FER_tau}.
On the other hand, when $\tau$ is sufficiently large, the improved gains tends to be saturated since only the most recent historical channels contribute to the majority of valuable features.

\subsection{Reliability-Latency Tradeoff}
\begin{figure}[t]
  \centering
  \includegraphics[width=3.6in,height=3.2in]{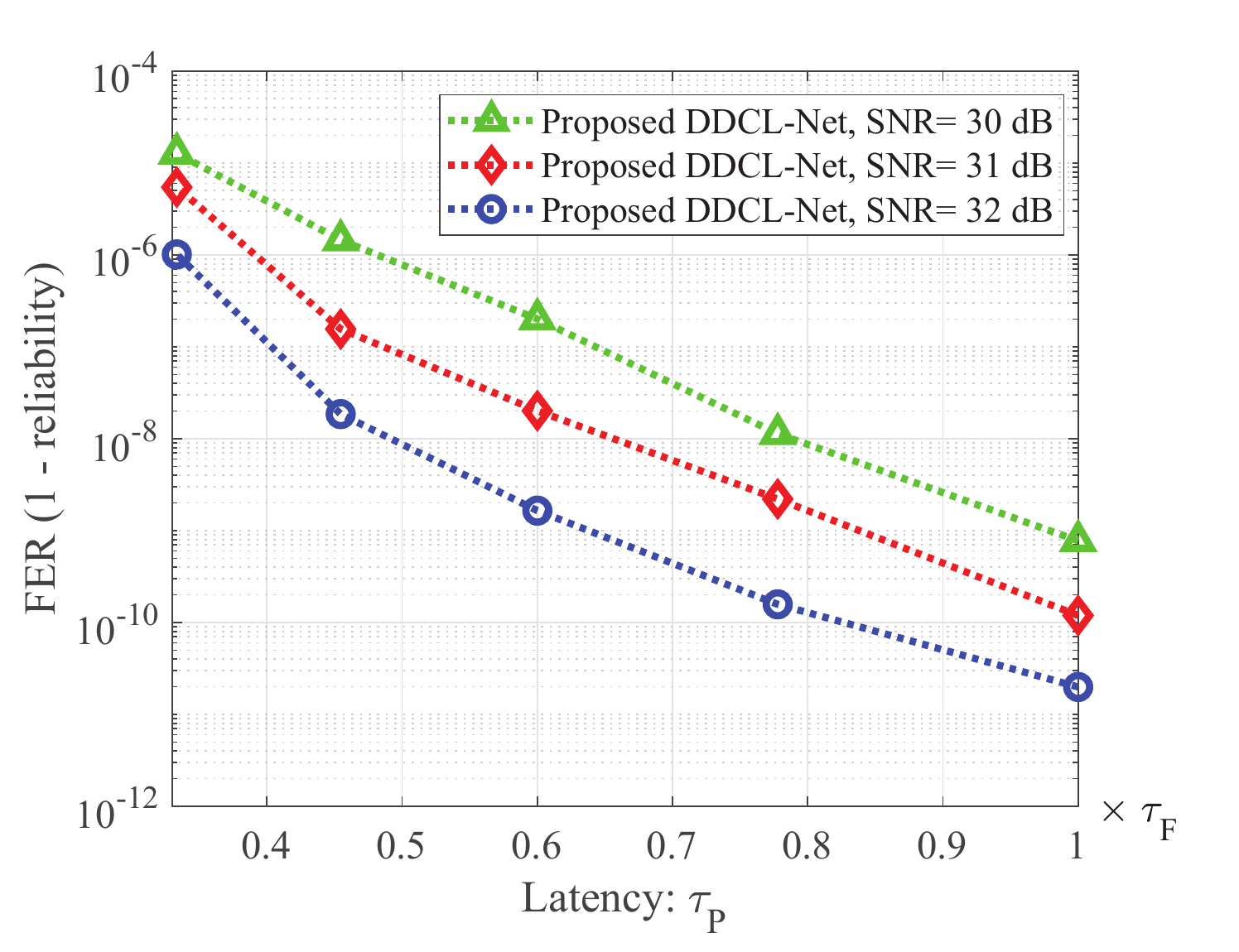}
  \caption{The reliability-latency tradeoff curves under $K=32$ and QPSK modulation.}\label{Fig:R_L_tradeoff}
\end{figure}

In this section, we study the reliability-latency tradeoff of our proposed method in the considered URLLC system.
Let $\gamma = \frac{K}{MN} \in (0,1]$ be the ratio of the original symbol length to the precoded symbol length.
Note that when $\gamma  < 1$, the original symbol length is smaller than the precoded symbol length, resulting in an additional latency. Since this latency is caused by the precoder operation, we call it precoder-caused latency, denoted by $\tau_{\mathrm{P}} = (\frac{1}{\gamma } - 1) \tau_{\mathrm{F}}$, where $\tau_{\mathrm{F}}$ is the duration of one frame.
The curves of reliability versus latency of the developed method in the considered OTFS-enabled URLLC system is presented in Fig. \ref{Fig:R_L_tradeoff}, where $\mathrm{FER}$ and $\tau_{\mathrm{P}}$ are adopted to characterize the system reliability and latency, respectively.
It can be observed that for different SNRs, FER decreases with the increase of latency, i.e., the reliability increases with the latency, which unveils the novel tradeoff phenomenon between reliability and latency.
The reason is that a long latency indicates a small value of $\gamma $, in this case, the proposed method can exploit DDCL-Net to intelligently design a longer sequence channel coding scheme \cite{tse2005fundamentals} to further improve the system reliability.
Therefore, by selecting appropriate value for $\gamma$, one can design a practical precoder to satisfy the desired point on the reliability-latency tradeoff curve, which provides more insight in balancing the reliability and latency performance for URLLC systems.

\section{Conclusion}
In this paper, we investigated the OTFS-enabled URLLC system and proposed a DL-based predictive precoder design scheme to further improve the system reliability.
In particular, we developed a predictive transmission protocol and proposed a universal DL-based predictive precoder design framework by exploiting the unsupervised learning mechanism.
To provide a realization of the developed framework, a DDCL-Net was designed to facilitate the predictive precoder design, where we leverage both the CNN and LSTM modules to exploit the spatial-temporal features from the  estimated historical channels to design precoder for the  subsequent time frame.
Finally, extensive simulation results were provided to verify the efficiency and effectiveness of the developed scheme in terms of FER performance, generalizability, and reliability-latency tradeoff.
Specifically, results demonstrated that the FER performance of the proposed method can approach that of the lower bound obtained by the perfect ICSI-aided scheme.

\bibliographystyle{ieeetr}

\setlength{\baselineskip}{10pt}

\bibliography{ReferenceSCI2}

\end{document}